\documentclass[sigconf,screen]{acmart} 
\setcopyright{acmlicensed}
\copyrightyear{2019}
\acmYear{2019}
\acmConference[ESEC/FSE '19]{Proceedings of the 27th ACM Joint European Software Engineering Conference and Symposium on the Foundations of Software Engineering}{August 26--30, 2019}{Tallinn, Estonia}
\acmBooktitle{Proceedings of the 27th ACM Joint European Software Engineering Conference and Symposium on the Foundations of Software Engineering (ESEC/FSE '19), August 26--30, 2019, Tallinn, Estonia}
\acmPrice{15.00}
\acmDOI{10.1145/3338906.3338909}
\acmISBN{978-1-4503-5572-8/19/08}

\usepackage{booktabs} 
\usepackage{colortbl}
\usepackage{amsmath}
\usepackage{algorithm}
\usepackage{algpseudocode}
\usepackage[normalem]{ulem} 
\usepackage{xspace}
\usepackage{multirow} 
\usepackage{balance} 
\usepackage{fancyvrb} 
\usepackage[export]{adjustbox} 
\usepackage[caption=false]{subfig}
\usepackage{url}
\usepackage{hyperref}
\usepackage{xcolor}

\usepackage{mdframed}
\mdfsetup{skipabove=0.5\topskip,skipbelow=0.5\topskip}

\usepackage{enumitem}
\setlist[itemize]{leftmargin=*,noitemsep,topsep=0pt}
\setlist[enumerate]{leftmargin=*}

\newenvironment{smalldescription}{
   \setlength{\topsep}{0pt}
   \setlength{\partopsep}{0pt}
   \setlength{\parskip}{0pt}
   \begin{description}[style=unboxed]
   \setlength{\leftmargin}{1in}
   \setlength{\parsep}{0pt}
   \setlength{\parskip}{0pt}
   \setlength{\itemsep}{0pt}
   }
   {\end{description}}

\usepackage{cleveref}
\crefformat{section}{\S#2#1#3}
\crefname{figure}{Figure}{Figures}
\crefname{table}{Table}{Tables}

\usepackage[disable]{todonotes}
\newcommand{\JD}[1]{\todo[color=yellow,inline]{JD:#1}}

\usepackage{soul}

\newcommand{\code}[1]{\texttt{{\small #1}}}

\newcommand{\ie}{\textit{i.e.,}\xspace}
\newcommand{\eg}{\textit{e.g.,}\xspace}
\newcommand{\etal}{\textit{et al.}\xspace}

\newcommand{\REDOS}{ReDoS\xspace}

\newcommand{\SurveyNumQuestions}{33\xspace}
\newcommand{\SurveyPercReuse}{94\%\xspace}
\newcommand{\SurveyPercReuseHalf}{50\%\xspace}
\newcommand{\SurveyFilteredOut}{253\xspace}

\newcommand{\SurveyNumValidResponses}{158\xspace}

\newcommand{\SurveyNumHundredLang}{34\xspace}
\newcommand{\SurveyDirectColleague}{51\xspace}
\newcommand{\SurveyNumSnowflakes}{25\xspace}

\newcommand{\SurveyNumInternetResponses}{73\xspace}
\newcommand{\SurveyNumAnonymousResponses}{9\xspace}

\newcommand{\SurveyNumNoLangMatterApproxPercent}{47\%\xspace}

\newcommand{\SurveyApproxPercReuseUseInternet}{90\%\xspace}

\newcommand{\SurveyApproxPercReuseUseCode}{90\%\xspace}

\newcommand{\WitnessesTableJavaScriptAnchors}{31}

\newcommand{\NumLanguagesInSample}{8\xspace}

\newcommand{\ApproxPercAllModulesWithReusedRegex}{20}
\newcommand{\ApproxPercAllModulesWithInternetRegex}{5}

\newcommand{\CorpusNModulesJavaScript}{24,997\xspace}
\newcommand{\CorpusNModulesJava}{24,986\xspace}
\newcommand{\CorpusNModulesPHP}{24,995\xspace}
\newcommand{\CorpusNModulesPython}{24,997\xspace}
\newcommand{\CorpusNModulesRuby}{24,999\xspace}
\newcommand{\CorpusNModulesGo}{24,997\xspace}
\newcommand{\CorpusNModulesPerl}{31,827\xspace}
\newcommand{\CorpusNModulesRust}{11,724\xspace}
\newcommand{\CorpusNModulesTotal}{193,524\xspace}

\newcommand{\CorpusUniqueRegexesJavaScript}{150,922\xspace}
\newcommand{\CorpusUniqueRegexesJava}{19,332\xspace}
\newcommand{\CorpusUniqueRegexesPHP}{44,237\xspace}
\newcommand{\CorpusUniqueRegexesPython}{43,896\xspace}
\newcommand{\CorpusUniqueRegexesRuby}{153,334\xspace}
\newcommand{\CorpusUniqueRegexesGo}{22,105\xspace}
\newcommand{\CorpusUniqueRegexesPerl}{142,777\xspace}
\newcommand{\CorpusUniqueRegexesRust}{2,025\xspace}
\newcommand{\CorpusSumRegexesAllLangsIncludingDups}{578,628\xspace}

\newcommand{\NumUniqueRegexesInCorpus}{537,806\xspace}

\newcommand{\RegexReuseMinLength}{15\xspace}

\newcommand{\SyntaxPercRegexWorkInEveryLang}{76}

\newcommand{\SyntaxPercRegexWorkNotRust}{88}

\newcommand{\SyntaxApproxPercRegexWorkInAtLeastSevenLangs}{92}

\newcommand{\SemanticExperimentFirstQuartileInputsPerRegex}{1,057\xspace}
\newcommand{\SemanticExperimentMedianInputsPerRegex}{2,410\xspace}

\newcommand{\SemanticExperimentThirdQuartileInputsPerRegex}{2,510\xspace}

\newcommand{\SemanticExperimentPercRegexesWithAnyWitnesses}{15.4}
\newcommand{\SemanticExperimentApproxPercRegexesWithAnyWitnesses}{15}
\newcommand{\SemanticExperimentPercRegexesWithMatchWitnesses}{8.1}
\newcommand{\SemanticExperimentPercRegexesWithSubstringWitnesses}{4.2}
\newcommand{\SemanticExperimentPercRegexesWithCaptureWitnesses}{7.5}
\newcommand{\SemanticExperimentNumRegexesWithMatchWitnesses}{43,417\xspace}
\newcommand{\SemanticExperimentNumRegexesWithSubstringWitnesses}{22,597\xspace}
\newcommand{\SemanticExperimentNumRegexesWithCaptureWitnesses}{40,457\xspace}

\newcommand{\SemanticExperimentNumRegexesWithWitnesses}{82,582}
\newcommand{\SemanticExperimentNumExplainedRegexes}{80,736}

\newcommand{\SemanticExperimentApproxPercExplainedRegexes}{98}

\newcommand{\SLExperimentSLDetectorTimeout}{60 seconds\xspace}
\newcommand{\SLExperimentSLDetectorMemoryLimit}{2 GB\xspace}
\newcommand{\SLExperimentEXPPumps}{100\xspace}
\newcommand{\SLExperimentPolyPumps}{100,000\xspace}
\newcommand{\SLExperimentSLThreshold}{10 seconds\xspace}

\newcommand{\SLExperimentNExpRegexesInPerl}{227\xspace}

\newcommand{\SLExperimentNaiveSpencerApproxNExp}{1,000\xspace}
\newcommand{\SLExperimentTotalNSLInRustAndGo}{6\xspace}

\newcommand{\SLExperimentNRegexesFoundByReScue}{1,421\xspace}

\newcommand{\SLExperimentApproxPercRegexesChangePerfBetweenAtLeastTwoLangs}{10}

\newcommand{\NumSnykDisclosedModules}{14,495\xspace}

\begin{document}

\VerbatimFootnotes 

\title[Why Aren't Regular Expressions a Lingua Franca?]{Why Aren't Regular Expressions a Lingua Franca? An Empirical Study on the Re-use and Portability of Regular Expressions}

\author{James C. Davis}
\email{davisjam@vt.edu}
\affiliation{%
  \institution{Virginia Tech, USA}
}

\author{Louis G. Michael IV}
\email{louism@vt.edu}
\affiliation{%
  \institution{Virginia Tech, USA}
}

\author{Christy A. Coghlan}
\authornote{Christy A. Coghlan is now employed by Google, Inc.}
\email{ccogs@vt.edu}
\affiliation{%
  \institution{Virginia Tech, USA}
}

\author{Francisco Servant}
\email{fservant@vt.edu}
\affiliation{%
  \institution{Virginia Tech, USA}
}

\author{Dongyoon Lee}
\email{dongyoon@vt.edu}
\affiliation{%
  \institution{Virginia Tech, USA}
}


%
%
%


\begin{abstract}
This paper explores the extent to which regular expressions (regexes) are portable across programming languages.
Many languages offer similar regex syntaxes, and it would be natural to assume that regexes can be ported across language boundaries.
But can regexes be copy/pasted across language boundaries while retaining their semantic and performance characteristics?

In our survey of \SurveyNumValidResponses professional software developers, most indicated that they re-use regexes across language boundaries and about half reported that they believe regexes are a universal language.
We experimentally evaluated the riskiness of this practice using a novel regex corpus --- \NumUniqueRegexesInCorpus regexes from \CorpusNModulesTotal projects written in JavaScript, Java, PHP, Python, Ruby, Go, Perl, and Rust.
Using our polyglot regex corpus, we explored the hitherto-unstudied \textbf{regex portability problems}: logic errors due to semantic differences, and security vulnerabilities due to performance differences.


We report that developers' belief in a regex \textit{lingua franca} is understandable but unfounded.
Though most regexes compile across language boundaries,
\SemanticExperimentApproxPercRegexesWithAnyWitnesses\% exhibit semantic differences across languages
and \SLExperimentApproxPercRegexesChangePerfBetweenAtLeastTwoLangs\% exhibit performance differences across languages.
We explained these differences using regex documentation,
and further illuminate our findings by investigating regex engine implementations.
Along the way we found bugs in the regex engines of JavaScript-V8, Python, Ruby, and Rust, and potential semantic and performance regex bugs in thousands of modules.



\end{abstract}



\maketitle

\vspace{-0.3cm}
\section{Introduction} \label{section:Introduction}


Regular expressions (regexes) are a core component of modern programming languages.
Regexes are commonly used for text processing and input sanitization~\cite{regexwikipedia}, appearing, for example, in an estimated 30-40\% of open-source Python and JavaScript projects~\cite{Chapman2016RegexUsageInPythonApps,Davis2018EcosystemREDOS}.
However, crafting a correct regex is difficult~\cite{Wang2019ExploringEvolution}, and developers may prefer to re-use 
an existing regex
than write it from scratch.
They might turn to regex repositories like RegExLib~\cite{RegexRepository_RegexLib,RegexRepository_Debuggex}; or to Stack Overflow, where ``regex'' is a popular tag~\cite{SO_MostPopularTags}; or to other software projects.
For example, a regex derived from the Node.js \code{path} module appears in more than 2,000 JavaScript projects~\cite{Davis2018EcosystemREDOS}.

Correctness and security are fundamental problems in software engineering in general, and for regexes in particular: \textit{re-using regexes can be risky}.
Like other code snippets~\cite{Yang2017SOGitHubSnippets}, regexes may flow into software from Internet forums or other software projects.
Unlike most code snippets, however, regexes can flow unchanged across language boundaries.
Programming languages have similar regex syntaxes, so re-used regexes may compile without modification.
However, surface-level syntactic compatibility can mask more subtle \textit{semantic} and \textit{performance} portability problems.
If regex semantics vary, then a regex will match different sets of strings across programming languages, resulting in logical errors.
If regex performance varies, a regex may have differing worst-case behavior, exposing service providers to security vulnerabilities~\cite{Crosby2003REDOS,Roichman2009ReDoS}.



Despite the widespread use of regexes in practice, the research literature is nearly silent about regex re-use and portability.
We know only anecdotally that some developers struggle with ``[regex] inconsistencies across [languages]''~\cite{Chapman2016RegexUsageInPythonApps}.
In this paper we explored the coupled concepts of cross-language regex re-use and regex portability using a mixed-methods approach.
First, we surveyed \SurveyNumValidResponses professional developers to better understand their regex practices (\cref{section:DeveloperSurvey}), and empirically corroborated the regex re-use practices they reported (\cref{section:CorroboratingSurvey}).
Then, we investigated the extent to which these practices may result in bugs.
We
  empirically measured semantic and performance portability problems,
  attempted to explain these problems using existing regex documentation,
  and explored regex engine implementations to illuminate our findings (\cref{section:Empirical}).
We are not the first to observe regex portability issues, but we are the first to provide evidence of the extent and impact of the phenomenon.

Our contributions are:
\begin{itemize}
    \item We describe the regex re-use practices of \SurveyNumValidResponses developers (\cref{section:DeveloperSurvey}).
    \item We present a first-of-its-kind \textit{polyglot regex corpus} consisting of \NumUniqueRegexesInCorpus unique regexes extracted from \CorpusNModulesTotal software projects written in \NumLanguagesInSample popular programming languages (\cref{section:RegexCorpus}).
    \item We empirically show that regex re-use is widespread, both within and across languages and from Internet sources (\cref{section:CorroboratingSurvey}).
    \item We identify and explain semantic and performance regex portability problems (\cref{section:Empirical}).
    We report that approximately \SemanticExperimentApproxPercRegexesWithAnyWitnesses\% of the regexes in our corpus have semantic portability problems, while \SLExperimentApproxPercRegexesChangePerfBetweenAtLeastTwoLangs\% have performance portability problems.
    Most of these problems could not be explained using existing regex documentation.
    \item We identify thousands of potential regex bugs in real software, as well as bugs in JavaScript-V8, Python, Ruby, and Rust (\cref{section:RegexBugs}).
\end{itemize}

\vspace{-0.1cm}
\section{Background} \label{section:Background}

\subsection{Regex Dialects} \label{section:Background-Regexes}


Most programming languages support regexes, providing developers with a concise means of describing a set of strings.
There has been no successful specification of regex syntax and semantics; Perl-Compatible Regular Expressions (PCRE)~\cite{PCRESpec} and POSIX Regular Expressions~\cite{POSIXStandard2018} have influenced but not standardized the various regex dialects that programming languages support, leading to manuals with phrases like ``these aspects...may not be fully portable''~\cite{man7regex}.
Anecdotally, inconsistent behavior has been reported even between different implementations of the same regex specification~\cite{Campeanu2009RegularRegexes}.

This lack of standardization may come as no surprise to developers familiar with regexes as a library feature rather than a language primitive.
But for the latest generation of developers, regexes have always been part of the programming language, and because the regex dialects are similar in syntax it would be natural for developers to assume that they are in fact a \textit{lingua franca}.
We report that many developers do make this assumption, and we explore the potential consequences by investigating the distinct semantic and performance characteristics of many regex dialects.





\vspace{-0.2cm}
\subsection{Regex Denial of Service (\REDOS)} \label{section:Background-REDOS}

Under the hood, programming languages implement a \textit{regex engine} to test candidate \textit{inputs} for membership in the language of a regex.
Most regex engines evaluate a regex match by simulating the behavior of an equivalent Finite Automaton (Deterministic or Non-) on the candidate string~\cite{Sipser2006AutomataTextbook}, but they vary widely in the particular algorithm used.
Despite the recommendations of automata theorists~\cite{Thompson1968LinearRegexAlgorithm,Cox2007RegexAlgorithms}, in most programming languages a regex match may require greater-than-linear time in the length of the regex and the input string.
Such a super-linear (SL) match may require polynomial or exponential time in the worst case~\cite{Crosby2003REDOS,Roichman2009ReDoS}.

SL regex behavior had long been considered an unlikely attack vector in practice, but in the past year this has begun to change.
Davis \etal~\cite{Davis2018NodeCure} and Staicu and Pradel~\cite{Staicu2018REDOS} identified Regular expression Denial of Service (\REDOS) as a major problem facing Node.js applications, and Davis \etal reported thousands of SL regexes affecting over 10,000 JavaScript projects~\cite{Davis2018EcosystemREDOS}.
Although it is known that SL regex behavior is possible in JavaScript, Python, and Java~\cite{Davis2018EcosystemREDOS,Weideman2016REDOSAmbiguity,Wustholz2017Rexploiter}, from a portability perspective we do not know the relative risk of \REDOS across different programming languages.
Cox~\cite{Cox2007RegexAlgorithms} has suggested that languages fall into two classes of performance, though he did not systematically support his claim.


\vspace{-0.1cm}
\subsection{Developer Practices Around Regexes} \label{section:Background-ExistingSurvey}

Despite the widespread use of regexes in practice~\cite{Chapman2016RegexUsageInPythonApps,Davis2018EcosystemREDOS}, surprisingly little is known about how software developers write and maintain them.
Recent studies have shed some light on
the typical languages developers encode in regexes~\cite{Chapman2016RegexUsageInPythonApps,Davis2018EcosystemREDOS},
the relative readability of different regex notations~\cite{Chapman2017RegexComprehension},
developer regex test practices~\cite{Wang2018RegexTestCoverage},
and developer regex maintenance practices~\cite{Davis2018EcosystemREDOS,Wang2019ExploringEvolution}.

Most of these works have focused on software artifacts rather than on developers' thought processes and day-to-day practices.
The only previous qualitative perspective on developers' approach to regex development is Chapman and Stolee's exploratory survey of 18 professional software developers employed by a single company~\cite{Chapman2016RegexUsageInPythonApps}.
They reported high-level regex practices like
the frequency with which those developers use regexes and
the tasks they use regexes for.
\section{Research Questions} \label{section:ResearchQuestions}

In this work we seek to better understand \textit{developer regex re-use practices} and understand the \textit{potential risks} they face.
First, we survey professional software developers to learn their regex perceptions and practices.
We then measure regex re-use practices in real software to corroborate the findings of our survey. 
Finally, we empirically evaluate the semantic and performance portability problems that may result from cross-language regex re-use practices, and explain differences across languages.
Our research questions:

\begin{smalldescription}
\item[Theme 1:] \textbf{Developer perspectives}
\item[RQ1:] Do developers re-use regexes?
\item[RQ2:] Where do developers re-use regexes from?
\item[RQ3:] Do developers believe regexes are a \textit{lingua franca}?
\end{smalldescription}

\begin{smalldescription}
\item[Theme 2:] \textbf{Measuring regex re-use}
\item[RQ4:] How commonly are regexes re-used from other software?
\item[RQ5:] How commonly are regexes re-used from Internet sources?
\end{smalldescription}


\begin{smalldescription}
\item[Theme 3:] \textbf{Empirical portability}
\item[RQ6:] \textit{Semantic portability:} When and why do regexes match different sets of strings in different programming languages?
\item[RQ7:] \textit{Performance portability:} When and why do regexes have different worst-case performance in different programming languages?
\end{smalldescription}



\section{Theme 1: Developer Perspectives} \label{section:DeveloperSurvey}

We surveyed developers to better understand
  regex re-use and portability issues
from their perspective.

\begin{mdframed}[backgroundcolor=black!10]
\textbf{Findings:}
(RQ1) \SurveyPercReuse of developers re-use regexes, \\ (RQ2) commonly from Stack Overflow and other code.
\\ (RQ3) \SurveyNumNoLangMatterApproxPercent of developers treat regexes like a \textit{lingua franca}.
\end{mdframed}

\subsection{Methodology} \label{section:DeveloperSurvey-Methodology}

\noindent\textbf{Survey content.}
We developed a \SurveyNumQuestions-question survey with a mix of closed and open-ended questions.
We asked participants about:
(1) the process they follow when writing regexes;
(2) their regex re-use practices;
and (3) what awareness they have of regex portability problems\footnote{Due to space limits we do not report all results.}.
We drafted our survey based on discussions with professional software developers, and followed best practices in survey design~\cite{Kitchenham2008PersonalSurveys,Siegmund2014MeasuringExperience}.
We refined the survey through internal iteration and two pilot rounds with graduate students.

\vspace{1.25mm}\noindent\textbf{Survey deployment.}
After obtaining approval from our institution's ethics board, we took a two-pronged approach to surveying professional developers.
First, following the snowball sampling methodology~\cite{Biernacki1981SnowballSampling,Sadler2010ResearchStrategy}, we asked developers of our acquaintance to take the survey and propagate it to their colleagues.
Second, to diversify our population, we posted the survey on popular Internet message boards frequented by software developers (HackerNews~\cite{HackerNews} and Reddit~\cite{Reddit} (\code{r/SampleSize}, \code{r/coding}, and \code{r/compsci}).
We compensated respondents with a \$5 Amazon gift card.


\vspace{1.25mm}\noindent\textbf{Filtering results.}
We received some invalid responses from users on the Internet message boards.
We manually inspected the first 100 responses to develop filters for validity.
We report on responses that
  took at least 5 minutes,
  were internally consistent,
  and
  gave a ``thoughtful'' answer to at least one of our open-ended questions.
This filtered out \SurveyFilteredOut responses, mostly from a spoofing campaign.


\subsection{Results} \label{section:DeveloperSurvey-Results}

\begin{figure}[thpb]
	\centering
	\includegraphics[width=1.0\columnwidth]{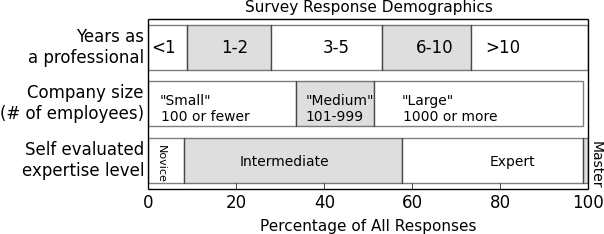}
	\caption{Our survey reached a diverse set of developers.
	}
	\label{figure:ProfDev}
\end{figure}

\noindent\textbf{Demographics.}
We received \SurveyNumValidResponses valid responses from professional software developers.
Our responses came from direct (\SurveyDirectColleague) and indirect (\SurveyNumSnowflakes) professional contacts, and Internet message boards (\SurveyNumInternetResponses), with no tracking information for \SurveyNumAnonymousResponses responses.
The median respondent has 3-5 years of professional experience, works at a medium-size company, and claims
intermediate regex skill\footnote{Regex skill was self-reported on a scale from novice to master, based on familiarity with increasingly complex regex features according to Friedl~\cite{friedl2006mastering}. ``Intermediate: For example, you have used more sophisticated features like non-greedy quantifiers (\verb|/a+?/|) and character classes (\verb</\d|\w|[abc]|[^\d]/<).''} (\cref{figure:ProfDev}). 

\begin{figure}[thpb]
	\centering
	\includegraphics[width=1.0\columnwidth]{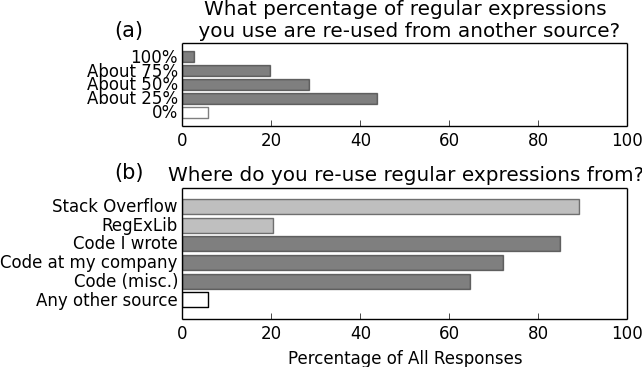}
	\caption{
	(a) When developers must use a regex, they frequently re-use them from another source.
	(b) Developers commonly re-use from the Internet and other software.
	}
	\label{figure:ReuseProcess}
\end{figure}

\vspace{1.25mm}\noindent\textbf{RQ1: Re-use Prevalence.}
Almost all (\SurveyPercReuse) of respondents indicated that they re-use regexes, with \SurveyPercReuseHalf indicating that they \textit{re-use} a regex at least half of the time that they \textit{use} a regex (\cref{figure:ReuseProcess} (a)). 
The most frequent reason to re-use a regex was to meet a common use case, \eg matching emails. 
This supports a previous hypothesis that developers may write regexes for a few common reasons \cite{Davis2018EcosystemREDOS}. 
Participants also mentioned time savings: ``A good programmer doesn't re-invent the wheel.''

\vspace{1.25mm}\noindent\textbf{RQ2: Re-use Sources.}
Developers most frequently said they re-use regexes from Stack Overflow, but they often re-use regexes from other code, including their own, a colleague's, or open-source projects (\cref{figure:ReuseProcess} (b)).
About \SurveyApproxPercReuseUseInternet of respondents reported re-using regexes from some Internet source, and about \SurveyApproxPercReuseUseCode reported re-using regexes from other code.

\begin{figure}[thpb]
	\centering
	\includegraphics[width=1.0\columnwidth]{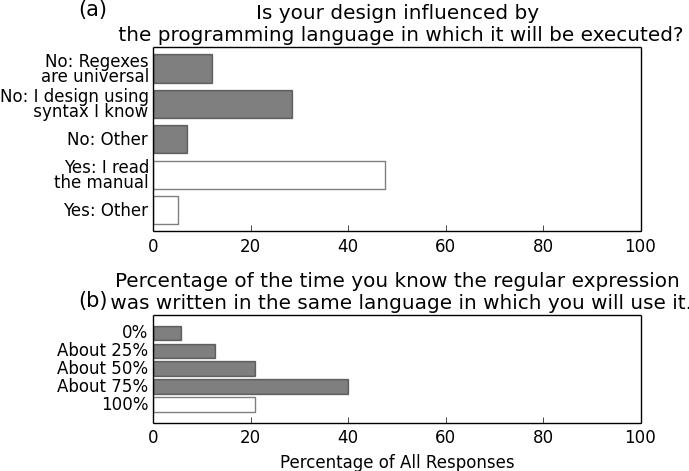}
	\caption{(a) Many developers design regexes without considering the programming language.
	(b) Developers' regex re-use decisions also imply belief in regex as a \textit{lingua franca}.
    }
	\label{figure:LangMatter}
	\vspace{-0.2cm}
\end{figure}

\vspace{1.25mm}\noindent\textbf{RQ3: Developer Perception of \textit{Lingua Franca}.}
We asked developers if their regex design process was influenced by language.
\cref{figure:LangMatter} (a) shows that \SurveyNumNoLangMatterApproxPercent of our respondents do not have a design process that is language specific.
And their actions match their beliefs:
as shown in~\cref{figure:LangMatter} (b), respondents frequently re-use regexes without being confident that they were written in the same language.
Only 21\% of respondents (\SurveyNumHundredLang/\SurveyNumValidResponses) were confident they never re-used across language boundaries.

\section{Polyglot Regex Corpus} \label{section:RegexCorpus}


In order to answer our remaining research questions we needed a \textit{polyglot regex corpus}: a set of regexes extracted from a large sample of software projects written in many programming languages.
The existing regex corpuses are small-scale~\cite{Chapman2016RegexUsageInPythonApps,Wustholz2017Rexploiter} or include only two programming languages~\cite{Davis2018EcosystemREDOS}.
Our corpus is neither, covering about 200,000 projects in \NumLanguagesInSample programming languages --- see \cref{table:Corpus}.

\vspace{-0.2cm}
\noindent\textbf{Languages.}
We are interested in studying common regex practices, and as a result we focus our attention on ``major'' programming languages defined by two conditions:
(1) The language has a large module ecosystem;
(2) The language is widely used by the open-source community.
We operationalized these concepts by consulting the ModuleCounts website~\cite{ModuleCounts} and the GitHub language popularity report~\cite{GitHubLanguagePopularityReport}.
We also considered Go, Perl, and Rust for scientific interest; Perl popularized the idea of regexes as a first-class language member, and Go and Rust are relatively new mainstream languages.

\begin{table}
\footnotesize
\centering
\caption{Our regex corpus was derived from software written in \NumLanguagesInSample programming languages. The first five languages are ranked by the most available libraries (ModuleCounts~\cite{ModuleCounts}) and popularity in open-source (GitHub).
We also studied Go, Perl, and Rust out of scientific interest.
The two final columns show the contribution to our corpus.
}
\begin{tabular}{lcccc}
\toprule
\textbf{Lang. (Registry)}  & \textbf{Libs.}   & \textbf{GH} & \textbf{\# mod. anal.}    & \textbf{Unique regexes (avg.)}           \\ \toprule
JavaScript\footnotemark \ (npm)               & 1                     & 1                 & \CorpusNModulesJavaScript & \CorpusUniqueRegexesJavaScript (6.0)  \\
Java (Maven)             & 2                     & 3                 & \CorpusNModulesJava       & \CorpusUniqueRegexesJava (0.8)         \\
PHP (Packagist)         & 3                     & 5                 & \CorpusNModulesPHP        & \CorpusUniqueRegexesPHP (1.2)          \\
Python (pypi)              & 4                     & 2                 & \CorpusNModulesPython     & \CorpusUniqueRegexesPython (1.8)       \\
Ruby (RubyGems)          & 5                     & 4                 & \CorpusNModulesRuby       & \CorpusUniqueRegexesRuby (6.1)         \\
\midrule
Go (Gopm)              & 9                     & 9                 & \CorpusNModulesGo         & \CorpusUniqueRegexesGo (0.9)           \\
Perl (CPAN)              & 7                     & ---               & \CorpusNModulesPerl (all)       & \CorpusUniqueRegexesPerl (4.5)         \\
Rust (Crates.io)         & 10                    & ---               & \CorpusNModulesRust (all)       & \CorpusUniqueRegexesRust (0.2)         \\
\midrule
 &  & \textit{Sum:} & \textit{\CorpusNModulesTotal} & \textit{\CorpusSumRegexesAllLangsIncludingDups}          \\
\bottomrule
\end{tabular}
\label{table:Corpus}
\vspace{-0.25cm}
\end{table}

\vspace{1.25mm}\noindent\textbf{Software projects.}
Within these languages, we chose to study the software modules published in each language's primary \textit{module registry} for two reasons.
First, it permits a relatively fair cross-language comparison, since we observe that many modules fill equivalent ecological niches, \eg logging or schema validation.
Second, we feel that modules are of greater general interest than applications.
Modules are published, maintained, and used by a mix of open-source and commercial software developers, and bugs and security vulnerabilities in modules have a significant ripple effect.

\footnotetext{We also extracted regexes from TypeScript source code, by transpiling it to JavaScript.}


Our goal was to analyze the most important modules in each language's primary module registry.
To have a uniform measure of importance across languages, we filtered each registry for the modules available on GitHub, sorted those by the number of stars, and analyzed the top 25,000 modules from each registry.
Borges and Valente recently showed that GitHub star count is a reasonable proxy for importance~\cite{Borges2018GitHubStars}.
Unsurprisingly, we found that the distribution of GitHub stars was similar for the modules in each language, and analyzing the top 25,000 modules typically captured all but the (very long) tail of 0-2 stars.
Perl and Rust had relatively few modules in their registries, and we analyzed all of their modules.
\vspace{1.25mm}\noindent\textbf{Regex extraction.}
Following~\cite{Davis2018EcosystemREDOS}, for each module we cloned the HEAD of its default branch from GitHub and extracted any statically-declared regexes.
We extracted regexes declared in regex evaluations as well as regexes compiled and stored in variables for later use.
In each module we extracted regexes only from source files in the language corresponding to the registry,
omitting regexes in places like build scripts written in another language.


\vspace{1.25mm}\noindent\textbf{Polyglot regex corpus.}
Our corpus contains \NumUniqueRegexesInCorpus unique regexes extracted from \CorpusNModulesTotal projects written in \NumLanguagesInSample programming languages.
Each language's contributions are listed in~\cref{table:Corpus}.
Average regex use varies widely by language, from 0.2 regexes per module (Rust) up to 6.1 regexes per module (Ruby).
The total unique regexes by language exceeds \NumUniqueRegexesInCorpus due to inter-language duplicates (\cref{section:CorroboratingSurvey}).

\section{Theme 2: Measuring regex re-use} \label{section:CorroboratingSurvey}

The developers in our survey indicated that they re-use regexes from other software and from Internet sources like Stack Overflow.
They also reported re-using regexes across language boundaries.
In this theme we corroborate their report by measuring the extent of regex re-use --- modules that use non-unique regexes.


\vspace{1.25mm}\noindent\textbf{Definition of re-use.}
To the best of our knowledge we are the first to attempt to measure regex re-use.
As a first approximation, in keeping with the phrasing in our survey (``copy/pasting regexes''), we label as re-use \textit{any pair of identical regexes} (string equality).
To eliminate trivially identical regexes like \verb|/\s/|, we conservatively require any matching regexes to be at least \RegexReuseMinLength characters long.
While it is possible that two developers might independently produce the same (longer) regex, this seems unlikely given that hundreds of distinct regexes have been reported even for ``simple'' languages like emails~\cite{Davis2018EcosystemREDOS}.
We do not consider less strict measures of regex equivalence like string ~\cite{Wang2019ExploringEvolution} or behavioral~\cite{Chapman2016RegexUsageInPythonApps} similarity, though such measures might better capture the ``Ship of Theseus'' approach to regex re-use described by some of our survey respondents.

\begin{mdframed}[backgroundcolor=black!10]
\textbf{Findings:}
(RQ4) Thousands of corpus modules (\ApproxPercAllModulesWithReusedRegex\%) share the same complex regexes, both within and across languages.
\\ (RQ5) \ApproxPercAllModulesWithInternetRegex\% of all corpus modules (about 10,000), primarily in JavaScript, use regexes from Stack Overflow and RegExLib.
\end{mdframed}

\subsection{RQ4: Re-use from Other Software} \label{section:RegexReuseSoftware}

How much intra-/inter-language regex re-use occurs in our corpus?

\subsubsection{Methodology} \label{section:RegexReuseSoftware-Methodology}

When developing our regex corpus (\cref{section:RegexCorpus}), we tracked the modules and registries in which each regex was found.
As noted above, we only consider as re-use candidates the regexes that are at least \RegexReuseMinLength characters long.
When such a regex appeared in multiple modules in the \textit{same} registry, we mark those modules as containing an intra-language duplicate.
When such a regex appeared in at least one module in \textit{different} registries, we mark those modules as containing an inter-language duplicate.
Note, then, that for a single duplicated regex we may mark several modules as containing intra-language duplicates and/or inter-language duplicates.

\subsubsection{Results} \label{section:RegexReuseSoftware-Results}

\begin{figure}[thpb]
	\centering
	\includegraphics[width=0.8\columnwidth]{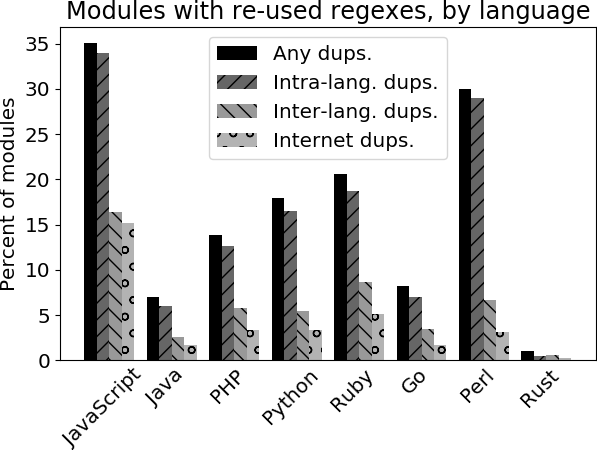}
	\caption{Empirical regex re-use practices, by language.}
	\label{figure:RegexReuse-All}
	\vspace{-0.4cm}
\end{figure}

The extent of intra- and inter-language regex re-use by modules is shown in~\cref{figure:RegexReuse-All} (second and third bars).
Developers re-use regexes in the modules in every language, some more than others.
In most languages, 10\% or more of the modules contain an intra-language duplicate, and inter-language duplicates are also common.
These duplicates are often due to ``popular'' regexes.
For example, we found the 16-character ``<email>:'' regex \verb|/[\w\-]+\@([^:]+):/| in 476 modules.


\subsection{RQ5: Re-use from Internet Sources} \label{section:InternetReuse}

The developers in our survey frequently indicated that they re-use regexes from one of two Internet sources: RegExLib~\cite{RegexRepository_RegexLib} and Stack Overflow~\cite{RegexRepository_StackOverflow}.
Next we use our corpus to corroborate their claims.

\subsubsection{Methodology} \label{section:InternetReuse-Methodology}

We extracted regexes from RegExLib and Stack Overflow, and then searched our corpus for matches.
In both of these relatively-unstructured Internet regex sources, the resulting set of ``regexes'' may include false positives; it is the intersection of our corpus and these sets that is of interest. 
An intersection is a case where a real regex from our corpus also appeared verbatim in one of these Internet sources.
Any module that contained one of these (15 characters or longer) Internet regexes was marked as containing an Internet duplicate.

\vspace{1.25mm}\noindent\textbf{RegExLib regexes.}
We obtained a reasonably complete dump of the RegExLib database by searching for ``all regexes''.

\vspace{1.25mm}\noindent\textbf{Stack Overflow regexes.}
For Stack Overflow, we relied on the ``regex'' tag to identify regexes.
Through manual analysis we found that questions and posts with the ``regex'' tag commonly denote regexes using code snippets.
Using all Stack Overflow posts as of September 2018\footnote{See \url{https://archive.org/download/stackexchange/stackoverflow.com-Posts.7z}.}, we found all questions tagged with ``regex'' as well as the answers to those questions and automatically extracted code snippets from those posts.
To filter, we then removed snippets that contained no regex-like characters based on Table 4 of~\cite{Chapman2016RegexUsageInPythonApps}.
\subsubsection{Results} \label{section:InternetReuse-Results}

Our findings are shown in~\cref{figure:RegexReuse-All} (fourth bar for each language).
Many of the modules in our corpus contain at least one Internet regex.
This practice is most common in JavaScript --- 15\% of npm modules contain an Internet regex.

\JD{James Donohue: ``Flow could be from code to SO''. We used to discuss this but dropped it. Worth at least a footnote in this section.}

\section{Theme 3: Empirical portability} \label{section:Empirical}

%

Having shown that developers re-use regexes across language boundaries,
now we experiment on our polyglot regex corpus to investigate the implications of copy/pasting a regex from one language to another.
First we consider semantic portability (\cref{section:Semantic}), then performance portability (\cref{section:SL}).


\vspace{1.25mm}\noindent\textbf{Experimental parameters.}
These experiments were performed on a 10-node cluster of server-class nodes.
We used the same base tools in both experiments: a tester for each of the \NumLanguagesInSample languages.
Each tester accepts a regex pattern and input and attempts a partial regex match.
\cref{table:Experiment-LanguageVersions} lists the language versions used in our tests.

When we compare a regex's behavior in a pair of languages, we use the subset of the regex corpus that is syntactically valid in that pair.
This simulates the regex re-use practices we identified.
Most comparisons are on the majority of the corpus --- \SyntaxPercRegexWorkInEveryLang\% of the corpus was valid in every language, and \SyntaxPercRegexWorkNotRust\% were valid in all but Rust.


\begin{table}
    \small
    \centering
    \caption{Summary of language versions and docs used in our experiments. Most are at the default for Ubuntu 16.04.}
    \begin{tabular}{ccc}
    \toprule
    \textbf{Language}   & \textbf{Version information}  & \textbf{Documentation} \\
    \toprule
    JavaScript          & Node.js v10.9.0 (V8 v6.8)     & \cite{JSRegexDocs1,JSRegexDocs2} \\
    Java                & Oracle JDK 8                  & \cite{JavaRegexDocs} \\
    PHP                 & PHP 7.4.0-dev (cli)           & \cite{PHPRegexDocs} \\
    Python              & Python 3.5.2                  & \cite{PythonRegexDocs1,PythonRegexDocs2} \\
    Ruby                & Ruby 2.3.1p112                & \cite{RubyRegexDocs} \\
    Go                  & Go v1.6.2                     & \cite{GoRegexDocs} \\
    Perl                & Perl v5.22.1                  & \cite{PerlRegexDocs1,PerlRegexDocs2,PerlRegexDocs3} \\
    Rust                & Rust v1.32.0 (nightly)        & \cite{RustRegexDocs} \\
    \bottomrule
    \end{tabular}
    \label{table:Experiment-LanguageVersions}
    \vspace{-0.25cm}
\end{table}

\begin{mdframed}[backgroundcolor=black!10]
\textbf{Findings:}
(RQ6) \SemanticExperimentApproxPercRegexesWithAnyWitnesses\% of regexes exhibit documented and undocumented \textit{semantic} differences.
(RQ7) \SLExperimentApproxPercRegexesChangePerfBetweenAtLeastTwoLangs\% of regexes exhibit \textit{performance} differences due to regex engine algorithms and optimizations.
\end{mdframed}

\vspace{-0.25cm}
\subsection{RQ6: Semantic Portability Problems} \label{section:Semantic}


When two languages express the same feature using different syntax, developers face a translation problem.
But when two languages exhibit different features (or behaviors) for the same syntax, developers must solve a semantic problem.
In this section we empirically study the semantic portability problems that developers may face.

\subsubsection{Methodology} \label{section:Semantic-Methodology}

To understand variations in regex semantics, we tested the behavior of each regex in our corpus on a variety of inputs in each language of interest.
Any inconsistent regex behavior across languages is something a developer would have to discover and address after re-using the regex.

\vspace{1.25mm}\noindent\textbf{Input generation.}
In search of an interesting set of inputs, we created an ensemble of five state-of-the-art regex input generators:
Rex~\cite{Veanes2010Rex},
MutRex~\cite{Arcaini2017MutRex},
EGRET~\cite{Larson2016EvilTestStrings},
ReScue~\cite{Shen2018ReScueGeneticRegexChecker}
and Brics~\cite{Moller2010Brics}.
These generators produce either matching strings (Rex, Brics) or both matching and mismatching strings (MutRex, EGRET, ReScue).
We used Rex, MutRex, and EGRET unchanged.
We modified ReScue to use the strings it explores in its search for SL inputs. 
We modified Brics to generate random input subsets, not infinitely many inputs.

We wanted these inputs to provide good regex automaton coverage.
Wang and Stolee showed that 100 Rex-generated inputs yield about 50\% regex coverage~\cite{Wang2018RegexTestCoverage}, so we requested 10,000 inputs from each input generator with a time limit of 10 seconds.
\cref{table:SemanticDifferencesSummary} summarizes the number of unique inputs generated for each regex.

\vspace{1.25mm}\noindent\textbf{Attempted match.}
For each regex, for each input, for each programming language that supported it, we tested for a match using partial-match semantics\footnote{We used default flags. As the \NumLanguagesInSample languages in our study support around 20 distinct regex flags, evaluating a meaningful subset of the flag combinations was infeasible.}.
On a match, we recorded (1) the substring that matched, and (2) the contents of capture groups.


\vspace{1.25mm}\noindent\textbf{Witnesses.}
Some pairs of languages may perfectly agree on the behavior of a regex on all of its inputs; others may not.
We refer to $(regex,input)$ pairs that produce different behavior in different programming languages as \textit{difference witnesses} between those languages, and distinguish between three disjoint types of witnesses:
\begin{enumerate}
\item \textit{Match witness}: Languages disagree on whether there is a match.
\item \textit{Substring witness}: Languages agree that there is a match but disagree about the matching substring.
\item \textit{Capture witness}: Languages agree on the match and the matching substring, but disagree about the division of the substring into any capture groups of the regex.
\end{enumerate}



\subsubsection{Results} \label{section:Semantic-Results}

\begin{figure}[thpb]
	\centering
	\includegraphics[width=1.0\columnwidth]{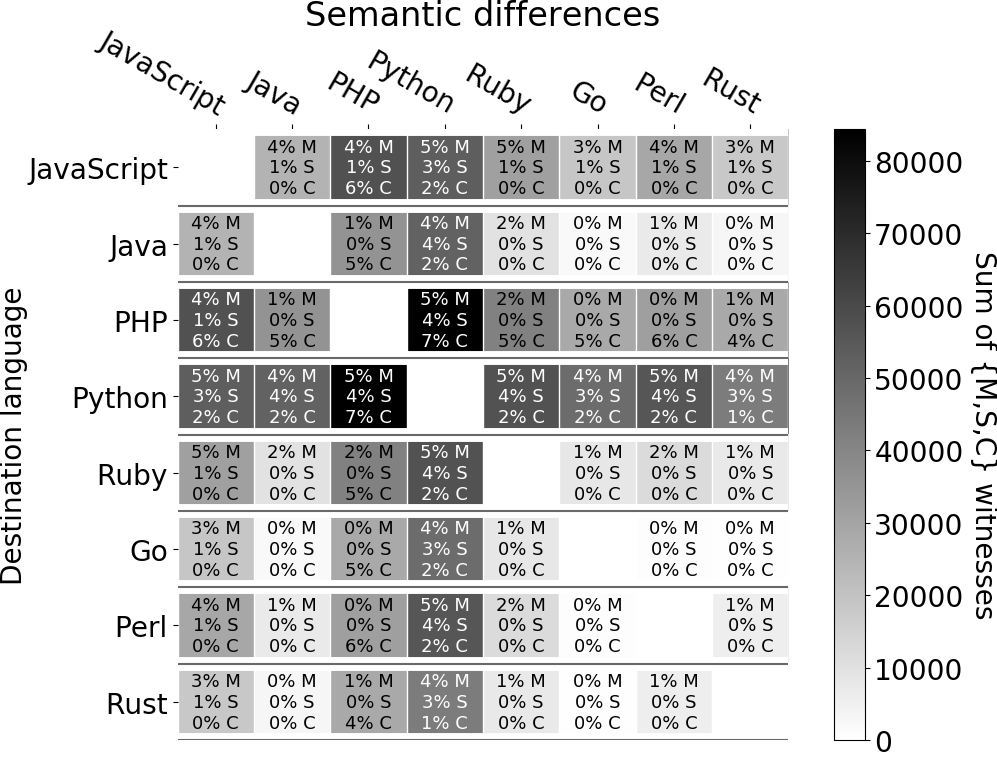}
	\caption{(Symmetric) Pairwise view of difference witnesses by language and type.
	The individual cells indicate the percent of the regex corpus with at least one (M)atch, (S)ubstring, and (C)apture witnesses in that language pair, and darker cells indicate that regexes more commonly have difference witnesses in that pair of languages.
	For example, Java, Go, and Rust generally agree on regex behavior.
	}
	\label{figure:SemanticDifferencesPortabilityHeatmap}
	\vspace{-0.3cm}
\end{figure}

\cref{table:SemanticDifferencesSummary} summarizes our results.
About \SemanticExperimentApproxPercRegexesWithAnyWitnesses\% of regexes participated in at least one difference witness, and among the language pairs we observed all three classes of witnesses.
In~\cref{table:SemanticDifferencesSummary} and~\cref{figure:SemanticDifferencesPortabilityHeatmap} we report the number of distinct regexes participating in the difference witnesses rather than the number of distinct witnesses themselves, because we expect that many of the witnessing inputs for a given regex are members of an equivalence class on which a difference manifests.

\begin{table}
\small
\centering
\caption{Statistics for semantic portability experiment.}
\begin{tabular}{cc}
\toprule
\textbf{Metric}                         & \textbf{Value} \\ \toprule
Percentile inputs per regex: 25$^{th}$-50$^{th}$-75$^{th}$ & \SemanticExperimentFirstQuartileInputsPerRegex-\SemanticExperimentMedianInputsPerRegex-\SemanticExperimentThirdQuartileInputsPerRegex \\
\midrule
Regexes with any difference witnesses              & \SemanticExperimentPercRegexesWithAnyWitnesses\% (\SemanticExperimentNumRegexesWithWitnesses) \\
\midrule
Regexes with any match witnesses        & \SemanticExperimentPercRegexesWithMatchWitnesses\% (\SemanticExperimentNumRegexesWithMatchWitnesses) \\
Regexes with any substring witnesses    & \SemanticExperimentPercRegexesWithSubstringWitnesses\% (\SemanticExperimentNumRegexesWithSubstringWitnesses) \\
Regexes with any capture witnesses      & \SemanticExperimentPercRegexesWithCaptureWitnesses\% (\SemanticExperimentNumRegexesWithCaptureWitnesses) \\
\bottomrule
\end{tabular}
\label{table:SemanticDifferencesSummary}
\vspace{-0.25cm}
\end{table}

A more detailed description of the semantic differences between languages is presented in~\cref{figure:SemanticDifferencesPortabilityHeatmap}.
The cells are colored proportional to the number of regexes that have any witness of a difference between that pair of languages.
The three numbers in the cell denote the percent\footnote{At the scale of our corpus, each percentage point represents about 5,300 regexes.} of regexes with match, substring, and capture witnesses for that pair of languages. 
As can be seen in~\cref{figure:SemanticDifferencesPortabilityHeatmap}: there are many language pairs with match witnesses; PHP and Python are the primary sources of substring witnesses; and PHP is the primary source of capture witnesses.

\subsubsection{Analysis} \label{section:Semantic-Analysis}

We developed an automatic tool, the \code{Cross Examiner}, to estimate the causes of the difference witnesses identified through our experiment.
We iteratively examined unclassified witnesses, referenced the regex documentation for the disagreeing languages (\cref{table:Experiment-LanguageVersions}), understood the reason for the different behaviors where documented, and encoded heuristics to classify witnesses as due to this behavior.
The causes we identified are summarized in~\cref{table:SemanticDifferencesCauses}.
Approximately \SemanticExperimentApproxPercExplainedRegexes\% (\SemanticExperimentNumExplainedRegexes/\SemanticExperimentNumRegexesWithWitnesses) of witnesses could be explained by one or more of these causes.

\cref{table:SemanticDifferencesCauses} differentiates the witnesses by type.
The first group of witnesses are cases where some languages support a feature that others do not.
In the second group, languages use the same syntax for different features.
The third group are cases where languages use the same syntax for the same features but exhibit different behavior.
The final group are bugs we identified, described below.

\vspace{1.25mm}\noindent\textbf{Documented semantics.}
We studied each language's regex documentation (\cref{table:Experiment-LanguageVersions}) to see if these witnesses could be easily explained.
Comparing the grey cells and boldfacing in~\cref{table:SemanticDifferencesCauses}, we note that more than half of the ``unusual'' behaviors were unspecified in that language's documentation.
\textbf{Testing, not reading the manual, is the only way for developers to learn these behaviors.}

\newcommand{\gc}{\cellcolor[gray]{0.9}}

\begin{table*}[t]
\footnotesize
\centering
\tabcolsep=0.08cm 
\caption{Difference witnesses identified during our semantic portability experiment.
Each row indicates a witness regex, the expected behavior(s), and each language's interpretation.
The first three groups describe different classes of valid but semantically distinct behavior.
The final group describes the bugs we found; E- means Engine, D- means Docs.
Boldface indicates potentially-surprising behavior (cf.~\cref{section:RegexBugs}).
``\code{-}'' indicates languages where a feature causes syntax errors.
The behavior in the grey cells was not specified in the documentation.
}
\begin{tabular}{llcccccccc}
\toprule
\textbf{Witness}                         & \textbf{Description} & \textbf{JavaScript} & \textbf{Java} & \textbf{PHP} & \textbf{Python} & \textbf{Ruby} & \textbf{Go} & \textbf{Perl} & \textbf{Rust} \\
\toprule

\multicolumn{8}{l}{\textbf{\textit{False friends 1: Regex notation describes a feature in one language and no feature in another.}}} \\

\verb|/\Qa\E/| & Quote directive ; ``QaE'' & \gc \textbf{``QaE''} & Quote & Quote & \textbf{``QaE''} & \gc \textbf{``QaE''} & Quote & Quote & \code{-} \\
\verb|/\G/| & Match assertion ; ``G'' & \gc \textbf{``G''} & Assertion & Assertion & \textbf{``G''} & Assertion & \code{-} & Assertion & \code{-} \\
\verb|/\Ab\Z/| & Anchors ; ``AbZ'' & \gc \textbf{``AbZ''} & Anchors & Anchors & Anchors & Anchors & \code{-} & Anchors & \code{-} \\
\verb|/a\z/| & End of line ; ``az'' & \gc \textbf{``az''} & EOL & EOL & \textbf{``az''} & EOL & EOL & EOL & EOL \\
\verb|/\K/|       & Match reset ; ``K'' & \gc \textbf{``K''} & \code{-} & Reset & \textbf{``K''} & Reset & \code{-} & Reset & \code{-} \\
\verb|/\e/|                & ESC ; ``e'' & \gc \textbf{``e''} & ESC & ESC & \textbf{``e''} & ESC & \code{-} & ESC & \code{-} \\
\verb|/\cC/|                & ctrl-C ; ``cC'' & ctrl-C & ctrl-C & ctrl-C & \textbf{``cC''} & ctrl-C & \code{-} & ctrl-C & \code{-} \\
\verb|/\x{41}/| & ``A'' (hex) ; ``x...x'' & \gc \textbf{``x...x''} & ``A'' & ``A'' & \code{-} & \code{-} & ``A'' & ``A'' & ``A'' \\
\verb|/(a)\g1/| & Backref notation ; ``ag1'' & \gc \textbf{``ag1''} & \code{-} & Backref & \textbf{``ag1''} & \gc \textbf{``ag1''} & \code{-} & Backref & \code{-} \\
\verb|/(a)\g<1>/| & Backref notation ; ``ag<1>'' & \gc \textbf{``ag<1>''} & \code{-} & Backref & \textbf{``ag<1>''} & Backref & \code{-} & \code{-} & \code{-} \\
\verb|/\p{N}/|   & Unicode digit ; ``p{N}'' & \gc \textbf{``p\{N\}}'' & 1 & 1 & \textbf{``p\{N\}''} & 1 & 1 & 1 & 1 \\
\verb|/\pN/|    & Unicode digit ; ``pN'' &  \gc \textbf{``pN''} & Digit & Digit & \textbf{``pN''} & \gc \textbf{``pN''} & Digit & Digit & Digit \\
\verb|/[[:digit:]]/| & Digit ; Custom Char. Class (CCC) & \textbf{CCC} & \textbf{CCC} & Digit & \textbf{CCC} & Digit & Digit & Digit & Digit \\

\midrule
\multicolumn{8}{l}{\textbf{\textit{False friends 2: The same regex notation describes different features.}}} \\

\verb|/^a/| & \verb|^|: Beginning of input or line & Input & Input & Input & Input & \textbf{Line} & Input & Input & Input \\
\verb|/a++/| & Possessive quantifier ; regular & \code{-} & Possessive & Possessive & \code{-} & Possessive & \code{-} & Possessive & \gc \textbf{Regular} \\
\verb|/(a)\1/| & Backref ; octal & Backref & Backref & Backref & Backref & Backref & \gc \code{-} & Backref & \textbf{Octal} \\
\verb|/\h/| & Horz. whitespace; Hex; ``h'' & \gc \textbf{``h''} & Whitespace & Whitespace & \textbf{``h''} & \textbf{Hex} & \code{-} & Whitespace & \code{-} \\

\midrule
\multicolumn{8}{l}{\textbf{\textit{Nuanced: The same regex notation describes the same feature, but engines exhibit subtly different behavior.}}} \\

\verb|/(a)(?<b>b)/| & Named and unnamed capture groups? & Both & Both & Both & \code{-} & \gc \textbf{Named only} & \code{-} & Both & \code{-} \\
\verb|/[]]/| & CCC of ``]'' ; empty CCC + ``]'' & \gc \textbf{Empty} & \gc ``]'' & \gc ``]'' & \gc ``]'' & \gc ``]'' & \gc ``]'' & \gc ``]'' & \gc ``]'' \\
\verb|/((a*)+)/| & Diff. capture of \textbackslash2 on ``aa'' & \gc \textbf{\textbackslash2: ``aa''} & \gc \textbackslash2: empty & \gc \textbackslash2: empty & \gc \textbackslash2: empty & \gc \textbackslash2: empty & \gc \textbf{\textbackslash2: ``aa''} & \gc \textbackslash2: empty & \gc \textbf{\textbackslash2: ``aa''} \\
\verb</((a)|(b))+/< & Diff. capture of \textbackslash2 on ``ab'' & \gc \textbf{Empty} & ``a'' & ``a'' & \gc ``a'' & \gc ``a'' & \gc ``a'' & \gc ``a'' & \gc ``a'' \\


\midrule
\midrule
\multicolumn{8}{l}{\textbf{\textit{Bugs we found in regex engines.}}} \\

E-Python: \verb</(ab|a)*?b/< & Diff. capture of \textbackslash1 on input: ``ab '' & ``a'' & ``a'' & ``a'' & \textbf{Empty} & ``a'' & ``a'' & ``a'' & ``a'' \\
E-Rust: \verb|/(aa$)?/| & Matched substring on ``aaz'' & Empty & Empty & Empty & Empty & Empty & Empty & Empty & \textbf{``aa''} \\
E-Rust: \verb|/(a)\d*\.?\d+\b/| & Matched substring on ``a0.0c '' & ``a0'' & ``a0'' & ``a0'' & ``a0'' & ``a0'' & ``a0'' & ``a0'' & \textbf{``a0.0''} \\
E-JavaScript: \textit{Complicated} & Input order matters? & \textbf{Yes} & No & No & No & No & No & No & No \\
D-OracleJava: \verb</$\s+/< & \$ matches before final \verb|\r|? & No & \textbf{Yes} & No & No & No & No & No & No \\
D-Ruby: \verb|/a{2}?/| & Lazy ``aa'' ; optional ``aa'' & Lazy & Lazy & Lazy & Lazy & \textbf{Optional} & Lazy & Lazy & Lazy \\
\bottomrule
\end{tabular}
\label{table:SemanticDifferencesCauses}
\vspace{-0.25cm}
\end{table*}


\subsubsection{Regex Engine Testing}

Though in this experiment we assumed that the regex engines were trustworthy, our methodology can be viewed as a mix of fuzz~\cite{chen2018systematic} and differential~\cite{mckeeman1998differential} testing.
Under a \textit{lingua franca} hypothesis, if languages disagree then at least one of them is wrong.
During our examination of difference witnesses, we identified five cases where one language disagreed with the others \textit{and} its behavior was inconsistent with the corresponding regex documentation.
We opened bug reports based on the behaviors briefly described in the third section of~\cref{table:SemanticDifferencesCauses}.
So far the bugs have been confirmed in V8-JavaScript, Python, Ruby, and Rust.
\subsection{RQ7: Performance Portability Problems} \label{section:SL}

Programming languages have distinct regex engines which may exhibit different performance characteristics.
A re-used regex might have worse worst-case performance in its new language than in its language of origin.
For example, software being ported from PHP to Node.js might develop Regular expression Denial of Service (\REDOS) vulnerabilities because regexes often have worse worst-case performance in Node.js.
In this experiment, we measure the frequency with which regexes have different worst-case performance characteristics in different programming languages.


\subsubsection{Methodology} \label{section:SL-Methodology}

We generally follow the methodology of Davis \etal~\cite{Davis2018EcosystemREDOS} and use their tools
In brief, for each regex we (1) query an ensemble of state-of-the-art super-linear regex detectors, and then (2) evaluate any predicted super-linear regex behaviors in each language of interest using partial-match semantics.

\vspace{1.25mm}\noindent\textbf{Experimental parameters.}
We allowed each of the detectors to evaluate a regex for up to \SLExperimentSLDetectorTimeout using no more than \SLExperimentSLDetectorMemoryLimit of memory.
If a detector predicted that a regex would be super-linear, we evaluated its proposed worst-case input in each of the \NumLanguagesInSample languages in our study using input strings intended to trigger exponential or polynomial behavior\footnote{We used \SLExperimentEXPPumps pumps for exponential and \SLExperimentPolyPumps pumps for polynomial.}.
If a regex match took more than \SLExperimentSLThreshold in some language, we marked it as super-linear.

\vspace{1.25mm}\noindent\textbf{Reducing false positives.}
We extended their methodology in two ways to reduce the number of false negatives (\ie SL regexes marked as linear-time).
First, we added Shen \etal's new dynamic SL regex detector~\cite{Shen2018ReScueGeneticRegexChecker}
to their ensemble (\cite{Rathnayake2014rxxr2,Weideman2016REDOSAmbiguity,Wustholz2017Rexploiter}).
More critically, we introduce a new technique that identifies both polynomial and exponential SL regexes that their detector ensemble would not detect.
The static detectors in their ensemble: (1) assume full-match semantics, and (2) do not scale well to regexes with large NFAs.
We combat these problems by querying detectors with the original regex as well as \textit{regex variants} that they can more readily analyze.


The first query variant addresses an \textit{unrealistic assumption} in the analysis performed by some of the detectors in the ensemble (\cite{Rathnayake2014rxxr2,Weideman2016REDOSAmbiguity,Wustholz2017Rexploiter}).
Although these detectors assume that the regex engine is using full-match semantics, regex engines generally default to partial-match semantics.
For example, some detectors predict linear behavior for \verb|/a+$/|, but it is quadratic in many languages when used with a partial-match API.
To address this assumption, we query the detector ensemble with an (anchored) full-match variant of unanchored regexes, \eg \verb|/^[\s\S]*?a+$/|.

The second query variant addresses \textit{inefficient implementations} in the detector ensemble.
Some of the detectors exceed our time limit
on regexes with large NFA representations.
For example, they time out on the (exponential) regex \verb|/(a{1,1000}){1,1000}$/| because its NFA explodes in size.
To account for this inefficiency, we query the detector ensemble with variants that replace bounded quantifiers with unbounded ones, \eg \verb|/(a+)+$/|.

These variants reduce the rate of false negatives without introducing false positives.
Although we query the detector ensemble on several variants, we always test any worst-case input on the original regex (dynamic validation).
The first variant may unmask polynomial regexes that would otherwise go undetected, and the second may identify both polynomial and exponential regexes.

\subsubsection{Results} \label{section:SL-Results}

\begin{figure}[t] 
	\centering
	\includegraphics[width=0.9\columnwidth]{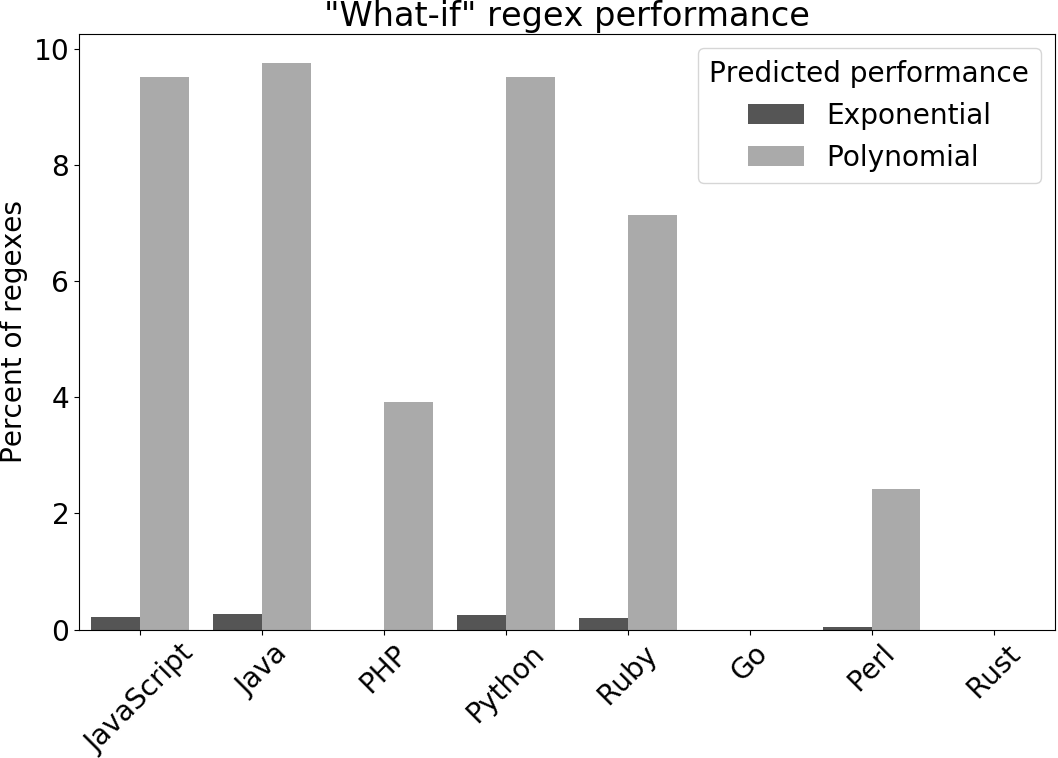}
	\caption{Proportion of SL regexes in each language.
	There are three distinct families of worst-case regex performance.
	We identified no regexes with exponential behavior in Go and Rust, and only \SLExperimentTotalNSLInRustAndGo regexes had polynomial behavior in those languages.
	Regexes with exponential behavior are rare in PHP and Perl (Perl -- \SLExperimentNExpRegexesInPerl; PHP -- 0), but polynomial behavior still occurs.
	In contrast, over \SLExperimentNaiveSpencerApproxNExp regexes have exponential behavior in Ruby, Java, JavaScript, and Python, and polynomial behavior is also more common in those languages.}
	\label{figure:SL-PerformanceByLanguage}
    \vspace{-0.3cm}
\end{figure}

\cref{figure:SL-PerformanceByLanguage} illustrates the extent to which the regexes in our polyglot corpus exhibit worst-case super-linear behavior in each of the \NumLanguagesInSample languages under study.

\cref{figure:SL-PerformanceByLanguage} indicates that SL regexes may be more common --- by up to an order of magnitude! --- than was previously reported~\cite{Davis2018EcosystemREDOS}.
The majority of the newly-discovered regexes were identified through our variant testing technique; as expected, the new detector by Shen \etal~\cite{Shen2018ReScueGeneticRegexChecker} identified only exponential regexes (\SLExperimentNRegexesFoundByReScue of them).
Our results agree with a small-scale estimate in Java~\cite{Wustholz2017Rexploiter}.
Although~\cref{figure:SL-PerformanceByLanguage} does not provide a direct comparison to~\cite{Davis2018EcosystemREDOS},
the same larger proportions occur when considering the subset of our corpus derived from JavaScript and Python (as theirs was).



\subsubsection{Analysis} \label{section:SL-Analysis}

\begin{figure}[thpb]
	\centering
	\includegraphics[width=1.0\columnwidth]{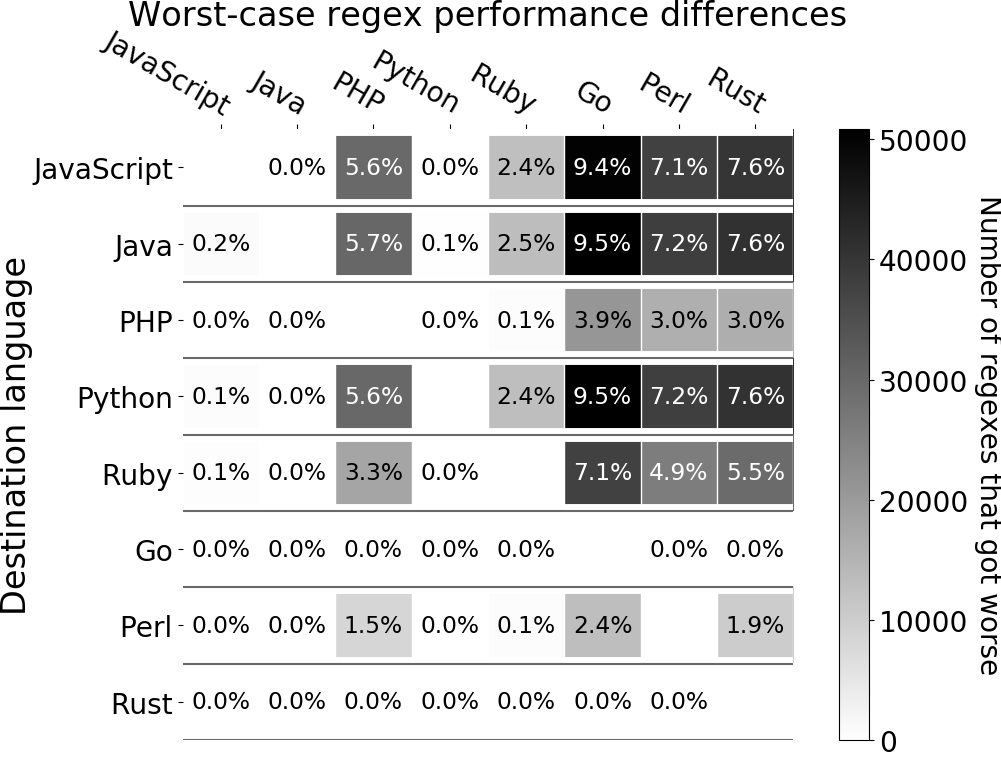}
	\caption{Pairwise view of regex performance differences.
	Cells are colored according to the number of regexes that exhibit worse behavior in the destination (row) than the hypothetical source (column).
	Darker rows are dangerous destinations; the individual cells contain the percent of the regexes supported in that language pair whose worst-case performance is worse in the destination.
	For example, regexes do not perform any worse in JavaScript than Java, but 8\% of regexes perform worse when moved from Rust to JavaScript.}
	\label{figure:SL-PortabilityHeatmap}
    \vspace{-0.4cm}
\end{figure}

The proportion of regexes that exhibit exponential and polynomial worst-case behavior varies widely by language.
The regex engines in these languages appear to fall into three families:
(1) \textit{Slow} (JavaScript, Java, Python, Ruby);
(2) \textit{Medium} (PHP, Perl); and
(3) \textit{Fast} (Go, Rust).
To clarify this taxonomy, \cref{figure:SL-PortabilityHeatmap} shows the frequency with which regexes exhibit \textit{worse} behavior in one of a pair of languages.
For example, we see that the {\textasciitilde}10\% of regexes that are super-linear in both Java and JavaScript (cf. \cref{figure:SL-PerformanceByLanguage}) are the \textit{same} regexes.
The worst-case performance of a regex generally worsens when moved between these families, but not within them.

In this section we explore the reasons behind these three families of regex performance.
We studied the language documentation and the implementation of these engines and identified a variety of mechanisms by which some regex engines fall prey to super-linear behavior and others avoid it.

\vspace{1.25mm}\noindent\textbf{Documented performance.}
We studied each language's regex documentation (\cref{table:Experiment-LanguageVersions}) to see whether its worst-case performance is discussed.
JavaScript, Java, and Python only provide tips on minor optimizations.
PHP and Ruby comment vaguely on worst-case performance: ``can take a long time to run''~\cite{PHPRegexDocs}.
The best documentation explicitly states worst-case expectations: linear (Rust and Go) or exponential (Perl).
Similar to the semantic behaviors in~\cref{table:SemanticDifferencesCauses}, in most languages these performance differences can only be identified through experimentation.

\vspace{1.25mm}\noindent\textbf{Under the hood.}
The primary distinction between these families is their core regex matching algorithms and varying support for super-linear regex features (\eg backreferences~\cite{Aho1990StringPatternAlgorithms}).
Go and Rust offer linear behavior because they primarily rely on Thompson's algorithm for linear-time regex evaluations~\cite{Thompson1968LinearRegexAlgorithm}, though in consequence they offer a limited set of regex features.
In contrast, the remaining 6 languages perform regex matches using some variant of Spencer's backtracking algorithm~\cite{Spencer1994RegexEngine}.
Thompson's algorithm is similar to a breadth-first traversal of the NFA graph, while Spencer's is analogous to a depth-first traversal.
Some implementations of the Spencer-style DFS may exhibit super-linear behavior due to redundant state visits, though there are also truly exponential (NP-complete) regexes with backreferences~\cite{Aho1990StringPatternAlgorithms,Berglund2017REWBR}.

Within the set of Spencer engines, though, there are distinct Medium and Slow Families.
In our experiments, exponential behavior was unusual in PHP and Perl, while it occurs at about the same rates in Java, JavaScript, Python, and Ruby.
Similarly, PHP and Perl have a lower incidence of polynomial behavior than do the other Spencer engines.
The differences between these two families can be attributed to a mix of \textit{defenses} and \textit{optimizations}.

To the best of our knowledge, PHP and Perl are the only Spencer engines in our study that have explicit defenses against exponential-time behavior.
Both languages rely on \textit{counters} to track the amount of work performed during a match, and if a regex evaluation exceeds a threshold it is terminated with an exception.
In experiments, we found that these counters are incremented such that exponential searches may trigger the threshold but poly-time searches will not.
Perl additionally maintains a \textit{cache} of visited states in order to short-circuit redundant paths through the NFA, permitting it to evaluate some searches in linear time that take polynomial or exponential time in other Spencer engines.

In addition to their exponential defenses, PHP and Perl both have optimizations that act as a safeguard against polynomial regex engine behavior.
For partial matches, some regex engines will try every possible starting offset in the string, trivially leading to polynomial behavior.
PHP and Perl have optimizations to prune these starting offsets, and these optimizations appear to reduce the incidence of polynomial behavior in those languages.
The relevant optimizations seem to be:
(1) skipping ahead to plausible starting points, and
(2) filtering out inputs that lack necessary substrings.

To the best of our knowledge, this is the first description of these real-world regex engine mechanisms in the scientific literature\footnote{Besides our description, we are only aware of descriptions of these defenses in discussion forum posts~\cite{PerlMonksResponseToCox} and the source code itself (\eg see line 7835 of~\cite{PerlRegexEngineSource_CacheCommentBlock}). These mechanisms are not described in the PHP and Perl documentation that we studied.}.
We hope our findings will inform the maintenance and development of regex engines that are less susceptible to super-linear behavior.

\vspace{1.25mm}\noindent\textbf{Three families, not two?}
In our experiments we were surprised to find three families of regex engine performance instead of the two previously described by Cox~\cite{Cox2007RegexAlgorithms}.
Perhaps based on Cox's analysis, others argued that exponential regex behavior in Java would translate to PHP~\cite{Shen2018ReScueGeneticRegexChecker}.
The defenses and optimizations we identified in PHP and Perl have previously gone unremarked.

\vspace{-0.15cm}
\section{Regex Bugs} \label{section:RegexBugs}


\textbf{Semantic bugs.}
Although developers may identify some semantic regex problems during testing, others may cause unexpected regex behavior in practice.
To estimate the frequency of semantic problems in practice, we developed linter-style tools to identify regexes that use features that are unavailable in their language (\cref{table:SemanticDifferencesCauses}).
For example, in JavaScript the anchor notation \verb|/\Ab\Z/| is interpreted literally as AbZ, but developers who use this notation in JavaScript projects probably intend anchors.
Among the JavaScript (npm) modules from which we derived our corpus, we identified \WitnessesTableJavaScriptAnchors\ modules that used this notation.
In total we identified hundreds of modules containing potential semantic regex bugs.
We have begun opening bug reports against these modules.

It is possible that these regexes were derived from copy/paste practices.
However, developers might introduce such bugs even when designing regexes from scratch, since they may design them based on a (supposed) regex \textit{lingua franca} that does not extend to the language in which they are developing (cf. \cref{figure:LangMatter}).

\vspace{1.25mm}\noindent\textbf{\REDOS regexes.}
The super-linear regexes we identified represent potential \REDOS vectors.
After filtering out regexes that appear in paths like \code{test} or \code{build}, we have initiated the responsible disclosure process to inform the developers of \NumSnykDisclosedModules modules about potential security vulnerabilities.

\vspace{-0.15cm}
\section{Discussion and Future Work} \label{section:Discussion}


\noindent\textbf{Considerations: software engineers.}
Our findings suggest that porting regexes across language boundaries, \eg from other software projects or from Stack Overflow, is a potentially risky activity.
Subtle semantic and performance issues can occur and should be considered by developers introducing regexes into their code.
Unfortunately, the largest developer communities are in the languages most vulnerable to \REDOS (cf. \cref{table:Corpus} and \cref{figure:SL-PerformanceByLanguage}).

We have released our many-language tools to help developers understand the possible risks of regexes.
Our tools can test the semantic and performance of regexes in many languages on many inputs.
We hope \cref{table:SemanticDifferencesCauses} will be a useful reference for developers.

\vspace{1.25mm}\noindent\textbf{Recommendations: programming language designers.}
We empathize with the developers we surveyed who expected regexes to behave consistently across programming languages.
We believe that regexes should truly be the \textit{lingua franca} many developers already believe them to be.
We suggest that having the fastest or most feature-rich engine is not worth the cost of regex portability problems.
Perhaps supported by researchers, programming language designers could agree on a universal regex specification and relieve software engineers of the burden of reconciling regexes across languages.
We acknowledge that diversity and competition sometimes improve outcomes for users, but regexes are a mature technology and unifying their behavior makes sense.

%

Each language's regex documentation currently focuses only on its own syntax and semantics.
We recommend that regex documentation additionally describe its deviations from external specification(s), \eg PCRE~\cite{PCRESpec} or PX-BRE~\cite{POSIXStandard2018}.
Explicitly discussing incompatibilities will inform developers of ``gotchas'', and it will have the indirect effect of reminding them that regexes are (currently) not a \textit{lingua franca}. 
Longer term, explicitly considering each language's divergence from specification(s) will help designers reach agreement on a next-generation universal regex specification.
Lastly, languages should document their worst-case regex performance.

We recommend that language designers in the ``Slow Family'' (JavaScript, Java, Python, Ruby) of regex engines adopt techniques from the ``Medium Family'' (PHP, Perl) to reduce the incidence of \REDOS vulnerabilities in these popular languages.


\vspace{1.25mm}\noindent\textbf{Corpus applications.}
Our polyglot regex corpus is a promising basis for further research.
Clearly developers search for regexes --- can we adapt semantic code search techniques~\cite{Ke2016SemanticCodeSearch} to the discovery of relevant regexes?
And do developers have different regex needs in different programming languages --- do these differences manifest in measurable ways, and should this affect the regex feature support or optimizations used in regex engines?

\vspace{1.25mm}\noindent\textbf{Regex tools and regex engines.}
Motivated by this work, we envision a regex ``universal translator'' to help developers port regexes between languages.
This task is complicated by incomplete regex specifications, different feature support in different programming languages, and performance variations.
As a starting point, van der Merwe \etal's work on regex transformations that preserve semantics but change performance seems promising~\cite{VanDerMerwe2017EvilRegexesHarmless}.

We believe that two directions for regex engine research are promising.
First, we accidentally identified bugs in four regex engines (\cref{section:Semantic-Analysis}).
Testing regex engines by refining regex semantics and applying model-checking techniques will improve the developer experience.
Second, we suggest that most \REDOS vulnerabilities can be solved at the regex engine level by refining and enhancing the optimizations already present in Perl and PHP (\cref{section:SL-Analysis}).
Care will be needed, however, to avoid changing regex semantics.



\vspace{1.25mm}\noindent\textbf{Other \textit{Lingua Francas}.}
What is the impact of other ``\textit{lingua franca} problems'' in software engineering?
For example, how do developers account for variations in SQL dialects, Markdown specifications, software compilers, and browser JavaScript support, and what are the consequences when they fail to do so?
\vspace{-0.10cm}
\section{Threats to Validity} \label{section:ThreatsToValidity}

\noindent\textbf{Internal validity.}
\textit{Survey.}
Our survey instrument has not been validated~\cite{Kitchenham2008PersonalSurveys}.
We assume the survey respondents who survived our ``bogus response'' filter replied in good faith.

\textit{Performance portability.}
Our results assume that the SL regex detector ensemble is effective. 
These detectors were designed with the naive Spencer-style regex engines in mind (``Slow family'') and might miss SL regexes in the Medium and Fast families.
For example, it is not clear whether the defenses of PHP and Perl are sound or simply effective against these detectors' inputs.


\vspace{1.25mm}\noindent\textbf{External validity.}
\textit{Regex corpus.}
Our methodology for procuring the regex corpus faces two threats.
First, our corpus is composed only of statically-declared regexes.
To generalize, we assume that either most regexes are statically declared, or that dynamically-declared regexes have similar properties.
Second, we only extract regexes from modules.
We do not know whether developers follow the same regex practices when writing regexes in modules and in applications, so our results may not generalize to applications.

\vspace{1.25mm}\noindent\textbf{Construct validity.}
\textit{Regex re-use.}
We took a simple approach to identifying regex re-use in our corpus: exact string matches for regexes at least 15 characters long.
We chose this threshold based on our assessment of regexes more or less likely to have been independently derived by multiple developers.
However, there may have been shorter re-used regexes, longer independently-derived regexes, and many regexes that were re-used with modifications to tailor them to specific use cases.

Our definition of re-use does not account for the possibility of wholesale file duplication,
which is not true regex re-use.
File duplication would only affect our intra-language regex re-use results.
\vspace{-0.10cm}
\section{Related Work} \label{section:RelatedWork}

\noindent\textbf{Empirical studies of regexes.}
The empirical study of regex use is a recent endeavor, with several lines of research.
Chapman and Stolee assessed the use of different regex features in Python~\cite{Chapman2016RegexUsageInPythonApps}.
Using that corpus, Chapman \etal assessed the relative understandability of regex synonyms to determine community preferences~\cite{Chapman2017RegexComprehension}.
Wang and Stolee reported that regexes are \textit{poorly unit tested} in Java applications~\cite{Wang2018RegexTestCoverage}, though this might be due to developer processes not captured by version control, \eg using regex checking tools~\cite{Larson2018AutomaticCheckingOfRegexes}.
Our polyglot regex techniques will enable generalizing some of these results to other programming languages.

\vspace{1.25mm}\noindent\textbf{Software re-use: other code.}
Software re-use is a prevalent practice in software engineering~\cite{Frakes2005,scacchi07fse,dubinsky2013exploratory,bauer2016comparing}.
Developers re-use code from their own or other projects~\cite{sim98}, introducing \emph{code clones}~\cite{roy09sci,gharehyazie18}.
Multiple studies estimate that more than 50\% of the code in GitHub is duplicated~\cite{Mockus2007CodeReuse,lopes17emse},
with similar ratios for Android applications~\cite{ruiz12}.

Whether code clones are good practice is a matter of debate.
Some researchers have pointed out the benefits of code clones~\cite{ossher11}, and found little difference between cloned and non-cloned code in qualities such as comprehensibility~\cite{saini18} and defect-proneness \cite{rahman12,sajnani14}.
But other studies have examined negative effects of cloning code~\cite{juergens09}, such as maintenance difficulties~\cite{fowler18} due to frequent~\cite{lozano07} but inconsistent~\cite{krinke07} changes.
As a result, a wide variety of techniques have been proposed to detect code clones, \eg~\cite{saini18fse,roy09sci,rattan13}.

To the best of our knowledge, our paper presents the first study of regex re-use and the problems that can arise from it.

\vspace{1.25mm}\noindent\textbf{Software re-use: Internet forums.}
Researchers have also studied software re-use from Internet forums.
Multiple studies found evidence of code flow from Stack Overflow to software repositories~\cite{Yang2017SOGitHubSnippets}, 
and found that code frequently flows, although
sometimes without respecting license terms~\cite{an17} or authorship attribution~\cite{baltes18emse}.
Researchers have also studied the interplay of developer contributions to both resources~\cite{vasilescu13}.

Given the prevalence of code snippets in Stack Overflow, multiple tools have been proposed to help developers re-use them, \eg to automatically generate comments \cite{wong13}, or to augment the IDE \cite{ponzanelli13,ponzanelli14}.
However, some problems have been identified with reusing code snippets from Stack Overflow, \eg quality~\cite{nasehi12} and usability~\cite{yang16}.
Furthermore, other studies have identified particular threats with code re-use from Stack Overflow, such as API misuse~\cite{Zhang2018AreSOCodeExamplesReliable}, security vulnerabilities~\cite{Fischer2017SOHarmfulSecSnippets,meng2018secure}, or unreadable code~\cite{treude17icsme}.

In this paper we found that Internet forums are also a popular source of regex re-use among developers, and we observed similar risks: feature mis-use and \REDOS vulnerabilities.

\vspace{1.25mm}\noindent\textbf{Migration.}
Researchers have long discussed the difficulties of code migration~\cite{Waters1988ProgramTranslation,Jacobson1991Old2OOP,Deursen1999LegacyRenovation,Terekhov2000LangConversion,Malton2001MigrationBarbell,Ray2013SemanticPortingProblems}.
As new technologies emerge, so do new migration tools,
\eg within~\cite{Barr2015Transplantation} and between languages~\cite{Samet1981Lisp2Lisp,Terekhov2001LanguageConversion,Martin2002C2Java,Mossienko2003Cobol2Java,ElRamly2006Java2CSharp,Zhong2010APIMapping,Phan2017MiningAPIUsage}
and frameworks~\cite{Lau2001DB2J2EE,Hassan2005MigratingWebFwork,Fan2016Android2iOS}.

Our work shows that regexes are (currently) not a \textit{lingua franca}, creating an opportunity for tools for regex migration.

\vspace{-0.1cm}
\section{Conclusion} \label{section:Conclusion}

Regexes are not a \textit{lingua franca}.
Although about \SyntaxApproxPercRegexWorkInAtLeastSevenLangs\% of regexes will compile in most programming languages, their apparent portability masks problems of correctness and performance.
We empirically investigated the extent and causes of these portability problems, offering the first empirical perspective on regex portability.
In the process we identified
hundreds of modules with potential semantic problems and
thousands with potential performance problems,
plus documentation and implementation errors in popular languages.

Unfortunately, but quite understandably, about half of the software developers we surveyed believe and act as though regexes are a \textit{lingua franca}.
We hope that this paper increases developer awareness of regex portability problems.
We also hope to motivate language designers toward regex standardization. 





\vspace{-0.15cm}
\section*{Reproducibility}

An artifact containing our survey instrument, regex corpus, and analyses is available at \url{https://doi.org/10.5281/zenodo.3257777}.

\vspace{-0.15cm}
\section*{Acknowledgments}

A. Kazerouni and R. Davis advised us on data analysis.
J. Donohue and E. Deram shared insights about developer re-use practices.
This work was supported in part by the National Science Foundation grant CNS-1814430.

\raggedbottom
\pagebreak

\balance

\bibliography{WebLinks.bib,LinguaFranca-static.bib,TmpRelated.bib,JamieBib-static.bib}


\begin{thebibliography}{111}


\ifx \showCODEN    \undefined \def \showCODEN     #1{\unskip}     \fi
\ifx \showDOI      \undefined \def \showDOI       #1{#1}\fi
\ifx \showISBNx    \undefined \def \showISBNx     #1{\unskip}     \fi
\ifx \showISBNxiii \undefined \def \showISBNxiii  #1{\unskip}     \fi
\ifx \showISSN     \undefined \def \showISSN      #1{\unskip}     \fi
\ifx \showLCCN     \undefined \def \showLCCN      #1{\unskip}     \fi
\ifx \shownote     \undefined \def \shownote      #1{#1}          \fi
\ifx \showarticletitle \undefined \def \showarticletitle #1{#1}   \fi
\ifx \showURL      \undefined \def \showURL       {\relax}        \fi
\providecommand\bibfield[2]{#2}
\providecommand\bibinfo[2]{#2}
\providecommand\natexlab[1]{#1}
\providecommand\showeprint[2][]{arXiv:#2}

\bibitem[\protect\citeauthoryear{??}{Hac}{[n.d.]}]%
        {HackerNews}
 \bibinfo{year}{[n.d.]}\natexlab{}.
\newblock \bibinfo{title}{Hacker News}.
\newblock \bibinfo{howpublished}{\url{https://news.ycombinator.com/}}.
\newblock


\bibitem[\protect\citeauthoryear{??}{Per}{[n.d.]}]%
        {PerlRegexDocs1}
 \bibinfo{year}{[n.d.]}\natexlab{}.
\newblock \bibinfo{title}{Perl Regular Expressions - Perl}.
\newblock
  \bibinfo{howpublished}{\url{https://perldoc.perl.org/5.22.0/perlre.html}}.
\newblock


\bibitem[\protect\citeauthoryear{??}{Red}{[n.d.]}]%
        {Reddit}
 \bibinfo{year}{[n.d.]}\natexlab{}.
\newblock \bibinfo{title}{Reddit}.
\newblock \bibinfo{howpublished}{\url{https://www.reddit.com/}}.
\newblock


\bibitem[\protect\citeauthoryear{??}{Reg}{[n.d.]a}]%
        {RegexRepository_RegexLib}
 \bibinfo{year}{[n.d.]}\natexlab{a}.
\newblock \bibinfo{title}{Regular Expression Library}.
\newblock
  \bibinfo{howpublished}{\url{https://web.archive.org/web/20180920164647/http://regexlib.com/}}.
\newblock


\bibitem[\protect\citeauthoryear{??}{Reg}{[n.d.]b}]%
        {RegexRepository_StackOverflow}
 \bibinfo{year}{[n.d.]}\natexlab{b}.
\newblock \bibinfo{title}{Stack Overflow - Regex tag}.
\newblock
  \bibinfo{howpublished}{\url{https://stackoverflow.com/questions/tagged/regex}}.
\newblock


\bibitem[\protect\citeauthoryear{??}{man}{2009}]%
        {man7regex}
 \bibinfo{year}{2009}\natexlab{}.
\newblock \bibinfo{title}{regex(7) - Linux manual page - POSIX.2 regular
  expressions}.
\newblock
  \bibinfo{howpublished}{\url{http://man7.org/linux/man-pages/man7/regex.7.html}}.
\newblock


\bibitem[\protect\citeauthoryear{??}{Reg}{2013}]%
        {RegexRepository_Debuggex}
 \bibinfo{year}{2013}\natexlab{}.
\newblock \bibinfo{title}{Debuggex: A Composable Regex Repository}.
\newblock
  \bibinfo{howpublished}{\url{https://web.archive.org/web/20170222084629/https://www.debuggex.com/blog/2013/a-composable-regex-repository/}}.
\newblock


\bibitem[\protect\citeauthoryear{??}{SO_}{2018}]%
        {SO_MostPopularTags}
 \bibinfo{year}{2018}\natexlab{}.
\newblock \bibinfo{title}{Tags -- Stack Overflow}.
\newblock
  \bibinfo{howpublished}{\url{https://web.archive.org/web/20180919183037/https://stackoverflow.com/tags?tab=popular}}.
\newblock


\bibitem[\protect\citeauthoryear{Aho}{Aho}{1990}]%
        {Aho1990StringPatternAlgorithms}
\bibfield{author}{\bibinfo{person}{Alfred~V Aho}.}
  \bibinfo{year}{1990}\natexlab{}.
\newblock \bibinfo{booktitle}{\emph{Algorithms for finding patterns in
  strings}}.
\newblock \bibinfo{publisher}{Elsevier}, Chapter~5, \bibinfo{pages}{255--300}.
\newblock


\bibitem[\protect\citeauthoryear{An, Mlouki, Khomh, and Antoniol}{An
  et~al\mbox{.}}{2017}]%
        {an17}
\bibfield{author}{\bibinfo{person}{Le An}, \bibinfo{person}{Ons Mlouki},
  \bibinfo{person}{Foutse Khomh}, {and} \bibinfo{person}{Giuliano Antoniol}.}
  \bibinfo{year}{2017}\natexlab{}.
\newblock \showarticletitle{Stack Overflow: A code laundering platform?}. In
  \bibinfo{booktitle}{\emph{International Conference on Software Analysis,
  Evolution and Reengineering (SANER)}}. IEEE.
\newblock


\bibitem[\protect\citeauthoryear{Arcaini, Gargantini, and Riccobene}{Arcaini
  et~al\mbox{.}}{2017}]%
        {Arcaini2017MutRex}
\bibfield{author}{\bibinfo{person}{Paolo Arcaini}, \bibinfo{person}{Angelo
  Gargantini}, {and} \bibinfo{person}{Elvinia Riccobene}.}
  \bibinfo{year}{2017}\natexlab{}.
\newblock \showarticletitle{{MutRex: A Mutation-Based Generator of Fault
  Detecting Strings for Regular Expressions}}. In
  \bibinfo{booktitle}{\emph{International Conference on Software Testing,
  Verification and Validation Workshops (ICSTW)}}.
\newblock
\showISBNx{9781509066766}


\bibitem[\protect\citeauthoryear{Baltes and Diehl}{Baltes and Diehl}{2018}]%
        {baltes18emse}
\bibfield{author}{\bibinfo{person}{Sebastian Baltes} {and}
  \bibinfo{person}{Stephan Diehl}.} \bibinfo{year}{2018}\natexlab{}.
\newblock \showarticletitle{Usage and attribution of Stack Overflow code
  snippets in GitHub projects}.
\newblock \bibinfo{journal}{\emph{Empirical Software Engineering}}
  (\bibinfo{year}{2018}), \bibinfo{pages}{1--37}.
\newblock


\bibitem[\protect\citeauthoryear{Barr, Harman, Jia, Marginean, and Petke}{Barr
  et~al\mbox{.}}{2015}]%
        {Barr2015Transplantation}
\bibfield{author}{\bibinfo{person}{Earl~T. Barr}, \bibinfo{person}{Mark
  Harman}, \bibinfo{person}{Yue Jia}, \bibinfo{person}{Alexandru Marginean},
  {and} \bibinfo{person}{Justyna Petke}.} \bibinfo{year}{2015}\natexlab{}.
\newblock \showarticletitle{{Automated Software Transplantation}}. In
  \bibinfo{booktitle}{\emph{International Symposium on Software Testing and
  Analysis (ISSTA)}}.
\newblock
\showISBNx{9781450336208}


\bibitem[\protect\citeauthoryear{Bauer et~al\mbox{.}}{Bauer
  et~al\mbox{.}}{2016}]%
        {bauer2016comparing}
\bibfield{author}{\bibinfo{person}{Veronika Bauer} {et~al\mbox{.}}}
  \bibinfo{year}{2016}\natexlab{}.
\newblock \showarticletitle{Comparing reuse practices in two large
  software-producing companies}.
\newblock \bibinfo{journal}{\emph{Journal of Systems and Software}}
  \bibinfo{volume}{117} (\bibinfo{year}{2016}), \bibinfo{pages}{545--582}.
\newblock


\bibitem[\protect\citeauthoryear{Berglund and Merwe}{Berglund and
  Merwe}{2017}]%
        {Berglund2017REWBR}
\bibfield{author}{\bibinfo{person}{Martin Berglund} {and}
  \bibinfo{person}{Brink Van~Der Merwe}.} \bibinfo{year}{2017}\natexlab{}.
\newblock \showarticletitle{{Regular Expressions with Backreferences}}. In
  \bibinfo{booktitle}{\emph{Prague Stringology}}. \bibinfo{pages}{30--41}.
\newblock
\showISBNx{9788001061930}


\bibitem[\protect\citeauthoryear{Biernacki and Waldorf}{Biernacki and
  Waldorf}{1981}]%
        {Biernacki1981SnowballSampling}
\bibfield{author}{\bibinfo{person}{Patrick Biernacki} {and}
  \bibinfo{person}{Dan Waldorf}.} \bibinfo{year}{1981}\natexlab{}.
\newblock \showarticletitle{{Snowball Sampling: Problems and Techniques of
  Chain Referral Sampling}}.
\newblock \bibinfo{journal}{\emph{Sociological Methods {\&} Research}}
  \bibinfo{volume}{10}, \bibinfo{number}{2} (\bibinfo{date}{11}
  \bibinfo{year}{1981}), \bibinfo{pages}{141--163}.
\newblock
\showISSN{0049-1241}


\bibitem[\protect\citeauthoryear{Borges and Tulio~Valente}{Borges and
  Tulio~Valente}{2018}]%
        {Borges2018GitHubStars}
\bibfield{author}{\bibinfo{person}{Hudson Borges} {and} \bibinfo{person}{Marco
  Tulio~Valente}.} \bibinfo{year}{2018}\natexlab{}.
\newblock \showarticletitle{{What's in a GitHub Star? Understanding Repository
  Starring Practices in a Social Coding Platform}}.
\newblock \bibinfo{journal}{\emph{Journal of Systems and Software}}
  \bibinfo{volume}{146} (\bibinfo{year}{2018}), \bibinfo{pages}{112--129}.
\newblock
\showISSN{01641212}
\urldef\tempurl%
\url{https://doi.org/10.1016/j.jss.2018.09.016}
\showDOI{\tempurl}


\bibitem[\protect\citeauthoryear{Britt and Laboratory}{Britt and
  Laboratory}{[n.d.]}]%
        {RubyRegexDocs}
\bibfield{author}{\bibinfo{person}{James Britt} {and}
  \bibinfo{person}{Neurogami~Secret Laboratory}.}
  \bibinfo{year}{[n.d.]}\natexlab{}.
\newblock \bibinfo{title}{Regexp - Ruby}.
\newblock
  \bibinfo{howpublished}{\url{https://ruby-doc.org/core-2.3.1/Regexp.html}}.
\newblock


\bibitem[\protect\citeauthoryear{C{\^{a}}mpeanu and Santean}{C{\^{a}}mpeanu and
  Santean}{2009}]%
        {Campeanu2009RegularRegexes}
\bibfield{author}{\bibinfo{person}{Cezar C{\^{a}}mpeanu} {and}
  \bibinfo{person}{Nicolae Santean}.} \bibinfo{year}{2009}\natexlab{}.
\newblock \showarticletitle{{On the intersection of regex languages with
  regular languages}}.
\newblock \bibinfo{journal}{\emph{Theoretical Computer Science}}
  \bibinfo{volume}{410}, \bibinfo{number}{24-25} (\bibinfo{year}{2009}),
  \bibinfo{pages}{2336--2344}.
\newblock
\showISSN{03043975}


\bibitem[\protect\citeauthoryear{Chapman and Stolee}{Chapman and
  Stolee}{2016}]%
        {Chapman2016RegexUsageInPythonApps}
\bibfield{author}{\bibinfo{person}{Carl Chapman} {and}
  \bibinfo{person}{Kathryn~T Stolee}.} \bibinfo{year}{2016}\natexlab{}.
\newblock \showarticletitle{{Exploring regular expression usage and context in
  Python}}.
\newblock \bibinfo{journal}{\emph{International Symposium on Software Testing
  and Analysis (ISSTA)}} (\bibinfo{year}{2016}).
\newblock
\showISBNx{9781450343909}


\bibitem[\protect\citeauthoryear{Chapman, Wang, and Stolee}{Chapman
  et~al\mbox{.}}{2017}]%
        {Chapman2017RegexComprehension}
\bibfield{author}{\bibinfo{person}{Carl Chapman}, \bibinfo{person}{Peipei
  Wang}, {and} \bibinfo{person}{Kathryn~T Stolee}.}
  \bibinfo{year}{2017}\natexlab{}.
\newblock \showarticletitle{{Exploring Regular Expression Comprehension}}. In
  \bibinfo{booktitle}{\emph{Automated Software Engineering (ASE)}}.
\newblock


\bibitem[\protect\citeauthoryear{Chen, Cui, Ma, Wu, Guo, and Liu}{Chen
  et~al\mbox{.}}{2018}]%
        {chen2018systematic}
\bibfield{author}{\bibinfo{person}{Chen Chen}, \bibinfo{person}{Baojiang Cui},
  \bibinfo{person}{Jinxin Ma}, \bibinfo{person}{Runpu Wu},
  \bibinfo{person}{Jianchao Guo}, {and} \bibinfo{person}{Wenqian Liu}.}
  \bibinfo{year}{2018}\natexlab{}.
\newblock \showarticletitle{A systematic review of fuzzing techniques}.
\newblock \bibinfo{journal}{\emph{Computers \& Security}}  \bibinfo{volume}{75}
  (\bibinfo{year}{2018}), \bibinfo{pages}{118--137}.
\newblock


\bibitem[\protect\citeauthoryear{Corp.}{Corp.}{[n.d.]}]%
        {JavaRegexDocs}
\bibfield{author}{\bibinfo{person}{Oracle Corp.}}
  \bibinfo{year}{[n.d.]}\natexlab{}.
\newblock \bibinfo{title}{Pattern - Java}.
\newblock
  \bibinfo{howpublished}{\url{https://docs.oracle.com/en/java/javase/11/docs/api/java.base/java/util/regex/Pattern.html}}.
\newblock


\bibitem[\protect\citeauthoryear{Cox}{Cox}{2007}]%
        {Cox2007RegexAlgorithms}
\bibfield{author}{\bibinfo{person}{Russ Cox}.} \bibinfo{year}{2007}\natexlab{}.
\newblock \bibinfo{title}{{Regular Expression Matching Can Be Simple And Fast
  (but is slow in Java, Perl, PHP, Python, Ruby, ...)}}.
\newblock
\newblock


\bibitem[\protect\citeauthoryear{Crosby}{Crosby}{2003}]%
        {Crosby2003REDOS}
\bibfield{author}{\bibinfo{person}{Scott Crosby}.}
  \bibinfo{year}{2003}\natexlab{}.
\newblock \showarticletitle{{Denial of service through regular expressions}}.
\newblock \bibinfo{journal}{\emph{USENIX Security work in progress report}}
  (\bibinfo{year}{2003}).
\newblock


\bibitem[\protect\citeauthoryear{Davis, Coghlan, Servant, and Lee}{Davis
  et~al\mbox{.}}{2018a}]%
        {Davis2018EcosystemREDOS}
\bibfield{author}{\bibinfo{person}{James~C Davis}, \bibinfo{person}{Christy~A
  Coghlan}, \bibinfo{person}{Francisco Servant}, {and}
  \bibinfo{person}{Dongyoon Lee}.} \bibinfo{year}{2018}\natexlab{a}.
\newblock \showarticletitle{{The Impact of Regular Expression Denial of Service
  (ReDoS) in Practice: an Empirical Study at the Ecosystem Scale}}. In
  \bibinfo{booktitle}{\emph{The ACM Joint European Software Engineering
  Conference and Symposium on the Foundations of Software Engineering
  (ESEC/FSE)}}.
\newblock


\bibitem[\protect\citeauthoryear{Davis, Williamson, and Lee}{Davis
  et~al\mbox{.}}{2018b}]%
        {Davis2018NodeCure}
\bibfield{author}{\bibinfo{person}{James~C Davis}, \bibinfo{person}{Eric~R
  Williamson}, {and} \bibinfo{person}{Dongyoon Lee}.}
  \bibinfo{year}{2018}\natexlab{b}.
\newblock \showarticletitle{{A Sense of Time for JavaScript and Node.js:
  First-Class Timeouts as a Cure for Event Handler Poisoning}}. In
  \bibinfo{booktitle}{\emph{USENIX Security Symposium (USENIX Security)}}.
\newblock


\bibitem[\protect\citeauthoryear{DeBill}{DeBill}{[n.d.]}]%
        {ModuleCounts}
\bibfield{author}{\bibinfo{person}{Erik DeBill}.}
  \bibinfo{year}{[n.d.]}\natexlab{}.
\newblock \bibinfo{title}{Module Counts}.
\newblock
  \bibinfo{howpublished}{\url{http://modulecounts-production.herokuapp.com/}}.
\newblock


\bibitem[\protect\citeauthoryear{Deursen, Klint, and Verhoef}{Deursen
  et~al\mbox{.}}{1999}]%
        {Deursen1999LegacyRenovation}
\bibfield{author}{\bibinfo{person}{Arie~van Deursen}, \bibinfo{person}{Paul
  Klint}, {and} \bibinfo{person}{Chris Verhoef}.}
  \bibinfo{year}{1999}\natexlab{}.
\newblock \showarticletitle{{Research Issues in the Renovation of Legacy
  Systems}}.
\newblock \bibinfo{journal}{\emph{Fundamental Approaches to Software
  Engineering}}  \bibinfo{volume}{1577} (\bibinfo{year}{1999}),
  \bibinfo{pages}{1--21}.
\newblock
\showISBNx{978-3-540-65718-7}
\showISSN{16113349}


\bibitem[\protect\citeauthoryear{Developers}{Developers}{[n.d.]}]%
        {RustRegexDocs}
\bibfield{author}{\bibinfo{person}{The Rust~Project Developers}.}
  \bibinfo{year}{[n.d.]}\natexlab{}.
\newblock \bibinfo{title}{regex - Rust}.
\newblock \bibinfo{howpublished}{\url{https://docs.rs/regex/1.1.0/regex/}}.
\newblock


\bibitem[\protect\citeauthoryear{Docs}{Docs}{[n.d.]a}]%
        {JSRegexDocs2}
\bibfield{author}{\bibinfo{person}{MDN~Web Docs}.}
  \bibinfo{year}{[n.d.]}\natexlab{a}.
\newblock \bibinfo{title}{RegExp - JavaScript}.
\newblock
  \bibinfo{howpublished}{\url{https://developer.mozilla.org/en-US/docs/Web/JavaScript/Reference/Global_Objects/RegExp}}.
\newblock


\bibitem[\protect\citeauthoryear{Docs}{Docs}{[n.d.]b}]%
        {JSRegexDocs1}
\bibfield{author}{\bibinfo{person}{MDN~Web Docs}.}
  \bibinfo{year}{[n.d.]}\natexlab{b}.
\newblock \bibinfo{title}{Regular Expressions - JavaScript}.
\newblock
  \bibinfo{howpublished}{\url{https://developer.mozilla.org/en-US/docs/Web/JavaScript/Guide/Regular_Expressions}}.
\newblock


\bibitem[\protect\citeauthoryear{Dubinsky, Rubin, Berger, Duszynski, Becker,
  and Czarnecki}{Dubinsky et~al\mbox{.}}{2013}]%
        {dubinsky2013exploratory}
\bibfield{author}{\bibinfo{person}{Yael Dubinsky}, \bibinfo{person}{Julia
  Rubin}, \bibinfo{person}{Thorsten Berger}, \bibinfo{person}{Slawomir
  Duszynski}, \bibinfo{person}{Martin Becker}, {and} \bibinfo{person}{Krzysztof
  Czarnecki}.} \bibinfo{year}{2013}\natexlab{}.
\newblock \showarticletitle{An exploratory study of cloning in industrial
  software product lines}. In \bibinfo{booktitle}{\emph{European Conference on
  Software Maintenance and Reengineering}}. IEEE.
\newblock


\bibitem[\protect\citeauthoryear{El-Ramly, Eltayeb, and Alla}{El-Ramly
  et~al\mbox{.}}{2006}]%
        {ElRamly2006Java2CSharp}
\bibfield{author}{\bibinfo{person}{M. El-Ramly}, \bibinfo{person}{R. Eltayeb},
  {and} \bibinfo{person}{H.A. Alla}.} \bibinfo{year}{2006}\natexlab{}.
\newblock \showarticletitle{{An Experiment in Automatic Conversion of Legacy
  Java Programs to C{\#}}}.
\newblock \bibinfo{journal}{\emph{IEEE International Conference on Computer
  Systems and Applications, 2006.}} \bibinfo{number}{March}
  (\bibinfo{year}{2006}), \bibinfo{pages}{1037--1045}.
\newblock
\showISBNx{1-4244-0211-5}


\bibitem[\protect\citeauthoryear{Fan and Wong}{Fan and Wong}{2016}]%
        {Fan2016Android2iOS}
\bibfield{author}{\bibinfo{person}{Xiaochao Fan} {and} \bibinfo{person}{Kenny
  Wong}.} \bibinfo{year}{2016}\natexlab{}.
\newblock \showarticletitle{{Migrating user interfaces in native mobile
  applications}}. In \bibinfo{booktitle}{\emph{International Workshop on Mobile
  Software Engineering and Systems (MOBILESoft)}}.
\newblock
\showISBNx{9781450341783}


\bibitem[\protect\citeauthoryear{Fischer, Bottinger, Xiao, Stransky, Acar,
  Backes, and Fahl}{Fischer et~al\mbox{.}}{2017}]%
        {Fischer2017SOHarmfulSecSnippets}
\bibfield{author}{\bibinfo{person}{Felix Fischer}, \bibinfo{person}{Konstantin
  Bottinger}, \bibinfo{person}{Huang Xiao}, \bibinfo{person}{Christian
  Stransky}, \bibinfo{person}{Yasemin Acar}, \bibinfo{person}{Michael Backes},
  {and} \bibinfo{person}{Sascha Fahl}.} \bibinfo{year}{2017}\natexlab{}.
\newblock \showarticletitle{{Stack Overflow Considered Harmful? the Impact of
  Copy{\&}Paste on Android Application Security}}. In
  \bibinfo{booktitle}{\emph{IEEE Symposium on Security and Privacy (IEEE
  S{\&}P)}}. \bibinfo{pages}{121--136}.
\newblock
\showISBNx{9781509055326}
\showISSN{10816011}


\bibitem[\protect\citeauthoryear{Foundation}{Foundation}{[n.d.]}]%
        {PythonRegexDocs1}
\bibfield{author}{\bibinfo{person}{Python~Software Foundation}.}
  \bibinfo{year}{[n.d.]}\natexlab{}.
\newblock \bibinfo{title}{re -- Regular expression operations - Python}.
\newblock
  \bibinfo{howpublished}{\url{https://docs.python.org/3.6/library/re.html}}.
\newblock


\bibitem[\protect\citeauthoryear{Fowler}{Fowler}{2018}]%
        {fowler18}
\bibfield{author}{\bibinfo{person}{Martin Fowler}.}
  \bibinfo{year}{2018}\natexlab{}.
\newblock \bibinfo{booktitle}{\emph{Refactoring: improving the design of
  existing code}}.
\newblock \bibinfo{publisher}{Addison-Wesley Professional}.
\newblock


\bibitem[\protect\citeauthoryear{{Frakes} and Kang}{{Frakes} and Kang}{2005}]%
        {Frakes2005}
\bibfield{author}{\bibinfo{person}{W.~B. {Frakes}} {and} \bibinfo{person}{Kyo
  Kang}.} \bibinfo{year}{2005}\natexlab{}.
\newblock \showarticletitle{Software reuse research: status and future}.
\newblock \bibinfo{journal}{\emph{IEEE Transactions on Software Engineering}}
  \bibinfo{volume}{31}, \bibinfo{number}{7} (\bibinfo{date}{July}
  \bibinfo{year}{2005}), \bibinfo{pages}{529--536}.
\newblock
\showISSN{0098-5589}


\bibitem[\protect\citeauthoryear{Friedl}{Friedl}{2006}]%
        {friedl2006mastering}
\bibfield{author}{\bibinfo{person}{Jeffrey~EF Friedl}.}
  \bibinfo{year}{2006}\natexlab{}.
\newblock \bibinfo{booktitle}{\emph{Mastering regular expressions}}.
\newblock \bibinfo{publisher}{" O'Reilly Media, Inc."}.
\newblock


\bibitem[\protect\citeauthoryear{Gharehyazie, Ray, Keshani, Zavosht,
  Heydarnoori, and Filkov}{Gharehyazie et~al\mbox{.}}{2018}]%
        {gharehyazie18}
\bibfield{author}{\bibinfo{person}{Mohammad Gharehyazie},
  \bibinfo{person}{Baishakhi Ray}, \bibinfo{person}{Mehdi Keshani},
  \bibinfo{person}{Masoumeh~Soleimani Zavosht}, \bibinfo{person}{Abbas
  Heydarnoori}, {and} \bibinfo{person}{Vladimir Filkov}.}
  \bibinfo{year}{2018}\natexlab{}.
\newblock \showarticletitle{Cross-project code clones in GitHub}.
\newblock \bibinfo{journal}{\emph{Empirical Software Engineering}}
  (\bibinfo{year}{2018}), \bibinfo{pages}{1--36}.
\newblock


\bibitem[\protect\citeauthoryear{GitHub}{GitHub}{2018}]%
        {GitHubLanguagePopularityReport}
\bibfield{author}{\bibinfo{person}{GitHub}.} \bibinfo{year}{2018}\natexlab{}.
\newblock \bibinfo{title}{The State of the Octoverse}.
\newblock \bibinfo{howpublished}{\url{https://octoverse.github.com/}}.
\newblock


\bibitem[\protect\citeauthoryear{Google}{Google}{[n.d.]}]%
        {GoRegexDocs}
\bibfield{author}{\bibinfo{person}{Google}.} \bibinfo{year}{[n.d.]}\natexlab{}.
\newblock \bibinfo{title}{regexp - Go}.
\newblock \bibinfo{howpublished}{\url{https://golang.org/pkg/regexp/}}.
\newblock


\bibitem[\protect\citeauthoryear{Group}{Group}{[n.d.]}]%
        {PHPRegexDocs}
\bibfield{author}{\bibinfo{person}{The~PHP Group}.}
  \bibinfo{year}{[n.d.]}\natexlab{}.
\newblock \bibinfo{title}{Regexp - PHP}.
\newblock
  \bibinfo{howpublished}{\url{http://php.net/manual/en/regexp.introduction.php}}.
\newblock


\bibitem[\protect\citeauthoryear{Hassan and Holt}{Hassan and Holt}{2005}]%
        {Hassan2005MigratingWebFwork}
\bibfield{author}{\bibinfo{person}{Ahmed~E. Hassan} {and}
  \bibinfo{person}{Richard~C. Holt}.} \bibinfo{year}{2005}\natexlab{}.
\newblock \showarticletitle{{A lightweight approach for migrating web
  frameworks}}.
\newblock \bibinfo{journal}{\emph{Information and Software Technology}}
  \bibinfo{volume}{47}, \bibinfo{number}{8} (\bibinfo{year}{2005}),
  \bibinfo{pages}{521--532}.
\newblock
\showISSN{09505849}


\bibitem[\protect\citeauthoryear{{Hazel, Philip}}{{Hazel, Philip}}{2018}]%
        {PCRESpec}
\bibfield{author}{\bibinfo{person}{{Hazel, Philip}}.}
  \bibinfo{year}{2018}\natexlab{}.
\newblock \bibinfo{title}{PCRE - Perl Compatible Regular Expressions}.
\newblock
  \bibinfo{howpublished}{\url{https://web.archive.org/web/20180919101106/https://www.pcre.org/}}.
\newblock


\bibitem[\protect\citeauthoryear{IEEE and Group}{IEEE and Group}{2018}]%
        {POSIXStandard2018}
\bibfield{author}{\bibinfo{person}{IEEE} {and} \bibinfo{person}{The~Open
  Group}.} \bibinfo{year}{2018}\natexlab{}.
\newblock \bibinfo{title}{The open group base specifications issue 7, 2018
  edition, ieee std 1003.1-2017}.
\newblock
\newblock


\bibitem[\protect\citeauthoryear{Jacobson and Lindstr{\"{o}}m}{Jacobson and
  Lindstr{\"{o}}m}{1991}]%
        {Jacobson1991Old2OOP}
\bibfield{author}{\bibinfo{person}{Ivar Jacobson} {and}
  \bibinfo{person}{Fredrik Lindstr{\"{o}}m}.} \bibinfo{year}{1991}\natexlab{}.
\newblock \showarticletitle{{Reengineering of old systems to an object-oriented
  architecture}}.
\newblock \bibinfo{journal}{\emph{ACM SIGPLAN Notices}} \bibinfo{volume}{26},
  \bibinfo{number}{11} (\bibinfo{year}{1991}), \bibinfo{pages}{340--350}.
\newblock
\showISBNx{0-201-55417-8}
\showISSN{03621340}


\bibitem[\protect\citeauthoryear{Juergens, Deissenboeck, Hummel, and
  Wagner}{Juergens et~al\mbox{.}}{2009}]%
        {juergens09}
\bibfield{author}{\bibinfo{person}{Elmar Juergens}, \bibinfo{person}{Florian
  Deissenboeck}, \bibinfo{person}{Benjamin Hummel}, {and}
  \bibinfo{person}{Stefan Wagner}.} \bibinfo{year}{2009}\natexlab{}.
\newblock \showarticletitle{Do code clones matter?}. In
  \bibinfo{booktitle}{\emph{International Conference on Software Engineering
  (ICSE)}}. IEEE.
\newblock


\bibitem[\protect\citeauthoryear{Ke, Stolee, Goues, and Brun}{Ke
  et~al\mbox{.}}{2016}]%
        {Ke2016SemanticCodeSearch}
\bibfield{author}{\bibinfo{person}{Yalin Ke}, \bibinfo{person}{Kathryn~T.
  Stolee}, \bibinfo{person}{Claire~Le Goues}, {and} \bibinfo{person}{Yuriy
  Brun}.} \bibinfo{year}{2016}\natexlab{}.
\newblock \showarticletitle{{Repairing programs with semantic code search}}. In
  \bibinfo{booktitle}{\emph{Automated Software Engineering (ASE)}}.
  \bibinfo{pages}{295--306}.
\newblock
\showISBNx{9781509000241}
\showISSN{0002-9262, 1476-6256}


\bibitem[\protect\citeauthoryear{Kitchenham and Pfleeger}{Kitchenham and
  Pfleeger}{2008}]%
        {Kitchenham2008PersonalSurveys}
\bibfield{author}{\bibinfo{person}{Barbara~A. Kitchenham} {and}
  \bibinfo{person}{Shari~L. Pfleeger}.} \bibinfo{year}{2008}\natexlab{}.
\newblock \showarticletitle{{Personal opinion surveys}}.
\newblock In \bibinfo{booktitle}{\emph{Guide to Advanced Empirical Software
  Engineering}}.
\newblock
\showISBNx{9781848000438}


\bibitem[\protect\citeauthoryear{Krinke}{Krinke}{2007}]%
        {krinke07}
\bibfield{author}{\bibinfo{person}{Jens Krinke}.}
  \bibinfo{year}{2007}\natexlab{}.
\newblock \showarticletitle{A study of consistent and inconsistent changes to
  code clones}. In \bibinfo{booktitle}{\emph{Working conference on reverse
  engineering (WCRE)}}. IEEE.
\newblock


\bibitem[\protect\citeauthoryear{Kuchling}{Kuchling}{[n.d.]}]%
        {PythonRegexDocs2}
\bibfield{author}{\bibinfo{person}{A.M. Kuchling}.}
  \bibinfo{year}{[n.d.]}\natexlab{}.
\newblock \bibinfo{title}{Regular Expression HOWTO - Python}.
\newblock
  \bibinfo{howpublished}{\url{https://docs.python.org/3.6/howto/regex.html}}.
\newblock


\bibitem[\protect\citeauthoryear{Kvale}{Kvale}{[n.d.]}]%
        {PerlRegexDocs3}
\bibfield{author}{\bibinfo{person}{Mark Kvale}.}
  \bibinfo{year}{[n.d.]}\natexlab{}.
\newblock \bibinfo{title}{Perl Regular Expressions Tutorial - Perl}.
\newblock
  \bibinfo{howpublished}{\url{https://perldoc.perl.org/5.22.0/perlretut.html}}.
\newblock


\bibitem[\protect\citeauthoryear{Larson}{Larson}{2018}]%
        {Larson2018AutomaticCheckingOfRegexes}
\bibfield{author}{\bibinfo{person}{Eric Larson}.}
  \bibinfo{year}{2018}\natexlab{}.
\newblock \showarticletitle{{Automatic Checking of Regular Expressions}}. In
  \bibinfo{booktitle}{\emph{Source Code Analysis and Manipulation (SCAM)}}.
\newblock


\bibitem[\protect\citeauthoryear{Larson and Kirk}{Larson and Kirk}{2016}]%
        {Larson2016EvilTestStrings}
\bibfield{author}{\bibinfo{person}{Eric Larson} {and} \bibinfo{person}{Anna
  Kirk}.} \bibinfo{year}{2016}\natexlab{}.
\newblock \showarticletitle{{Generating Evil Test Strings for Regular
  Expressions}}. In \bibinfo{booktitle}{\emph{International Conference on
  Software Testing, Verification and Validation (ICST)}}.
\newblock
\showISBNx{9781509018260}


\bibitem[\protect\citeauthoryear{Lau, Lu, Hedges, and Xing}{Lau
  et~al\mbox{.}}{2001}]%
        {Lau2001DB2J2EE}
\bibfield{author}{\bibinfo{person}{Terry Lau}, \bibinfo{person}{Jianguo Lu},
  \bibinfo{person}{Erik Hedges}, {and} \bibinfo{person}{Emily Xing}.}
  \bibinfo{year}{2001}\natexlab{}.
\newblock \showarticletitle{{Migrating E-commerce Database Applications to an
  Enterprise Java Environment}}. In \bibinfo{booktitle}{\emph{Conference of the
  Centre for Advanced Studies on Collaborative Research}}.
\newblock


\bibitem[\protect\citeauthoryear{Lopes, Maj, Martins, Saini, Yang, Zitny,
  Sajnani, and Vitek}{Lopes et~al\mbox{.}}{2017}]%
        {lopes17emse}
\bibfield{author}{\bibinfo{person}{Cristina~V Lopes}, \bibinfo{person}{Petr
  Maj}, \bibinfo{person}{Pedro Martins}, \bibinfo{person}{Vaibhav Saini},
  \bibinfo{person}{Di Yang}, \bibinfo{person}{Jakub Zitny},
  \bibinfo{person}{Hitesh Sajnani}, {and} \bibinfo{person}{Jan Vitek}.}
  \bibinfo{year}{2017}\natexlab{}.
\newblock \showarticletitle{D{\'e}j{\`a}Vu: a map of code duplicates on
  GitHub}.
\newblock \bibinfo{journal}{\emph{Proceedings of the ACM on Programming
  Languages (OOPSLA)}}.
\newblock


\bibitem[\protect\citeauthoryear{Lozano, Wermelinger, and Nuseibeh}{Lozano
  et~al\mbox{.}}{2007}]%
        {lozano07}
\bibfield{author}{\bibinfo{person}{Angela Lozano}, \bibinfo{person}{Michel
  Wermelinger}, {and} \bibinfo{person}{Bashar Nuseibeh}.}
  \bibinfo{year}{2007}\natexlab{}.
\newblock \showarticletitle{Evaluating the harmfulness of cloning: A change
  based experiment}. In \bibinfo{booktitle}{\emph{Mining Software Repositories
  (MSR)}}. IEEE.
\newblock


\bibitem[\protect\citeauthoryear{Malton}{Malton}{2001}]%
        {Malton2001MigrationBarbell}
\bibfield{author}{\bibinfo{person}{Andrew~J Malton}.}
  \bibinfo{year}{2001}\natexlab{}.
\newblock \showarticletitle{{The Software Migration Barbell}}.
\newblock \bibinfo{journal}{\emph{Proceedings of the ASERC Workshop on Software
  Architecture}} (\bibinfo{year}{2001}).
\newblock


\bibitem[\protect\citeauthoryear{Martin and Muller}{Martin and Muller}{2002}]%
        {Martin2002C2Java}
\bibfield{author}{\bibinfo{person}{J. Martin} {and} \bibinfo{person}{H.a.
  Muller}.} \bibinfo{year}{2002}\natexlab{}.
\newblock \showarticletitle{{C to Java migration experiences}}.
\newblock \bibinfo{journal}{\emph{Proceedings of the Sixth European Conference
  on Software Maintenance and Reengineering}} (\bibinfo{year}{2002}),
  \bibinfo{pages}{143--153}.
\newblock
\showISBNx{0-7695-1438-3}


\bibitem[\protect\citeauthoryear{McKeeman}{McKeeman}{1998}]%
        {mckeeman1998differential}
\bibfield{author}{\bibinfo{person}{William~M McKeeman}.}
  \bibinfo{year}{1998}\natexlab{}.
\newblock \showarticletitle{Differential testing for software}.
\newblock \bibinfo{journal}{\emph{Digital Technical Journal}}
  \bibinfo{volume}{10}, \bibinfo{number}{1} (\bibinfo{year}{1998}),
  \bibinfo{pages}{100--107}.
\newblock


\bibitem[\protect\citeauthoryear{Meng, Nagy, Yao, Zhuang, and
  Arango-Argoty}{Meng et~al\mbox{.}}{2018}]%
        {meng2018secure}
\bibfield{author}{\bibinfo{person}{Na Meng}, \bibinfo{person}{Stefan Nagy},
  \bibinfo{person}{Danfeng Yao}, \bibinfo{person}{Wenjie Zhuang}, {and}
  \bibinfo{person}{Gustavo Arango-Argoty}.} \bibinfo{year}{2018}\natexlab{}.
\newblock \showarticletitle{Secure coding practices in java: Challenges and
  vulnerabilities}. In \bibinfo{booktitle}{\emph{International Conference on
  Software Engineering (ICSE)}}. IEEE.
\newblock


\bibitem[\protect\citeauthoryear{Mockus}{Mockus}{2007}]%
        {Mockus2007CodeReuse}
\bibfield{author}{\bibinfo{person}{Audris Mockus}.}
  \bibinfo{year}{2007}\natexlab{}.
\newblock \showarticletitle{{Large-scale code reuse in open source software}}.
  In \bibinfo{booktitle}{\emph{First International Workshop on Emerging Trends
  in FLOSS Research and Development, FLOSS'07}}.
\newblock
\showISBNx{0769529615}


\bibitem[\protect\citeauthoryear{M{\o}ller}{M{\o}ller}{2010}]%
        {Moller2010Brics}
\bibfield{author}{\bibinfo{person}{Anders M{\o}ller}.}
  \bibinfo{year}{2010}\natexlab{}.
\newblock \bibinfo{title}{dk. brics. automaton--finite-state automata and
  regular expressions for Java, 2010}.
\newblock
\newblock


\bibitem[\protect\citeauthoryear{Mossienko}{Mossienko}{2003}]%
        {Mossienko2003Cobol2Java}
\bibfield{author}{\bibinfo{person}{M. Mossienko}.}
  \bibinfo{year}{2003}\natexlab{}.
\newblock \showarticletitle{{Automated Cobol to Java recycling}}. In
  \bibinfo{booktitle}{\emph{Conference on Software Maintenance and
  Reengineering (CSMR)}}, Vol.~\bibinfo{volume}{7}. \bibinfo{publisher}{IEEE},
  \bibinfo{pages}{40--50}.
\newblock
\showISBNx{0769519024}
\showISSN{15345351}


\bibitem[\protect\citeauthoryear{Nasehi, Sillito, Maurer, and Burns}{Nasehi
  et~al\mbox{.}}{2012}]%
        {nasehi12}
\bibfield{author}{\bibinfo{person}{Seyed~Mehdi Nasehi},
  \bibinfo{person}{Jonathan Sillito}, \bibinfo{person}{Frank Maurer}, {and}
  \bibinfo{person}{Chris Burns}.} \bibinfo{year}{2012}\natexlab{}.
\newblock \showarticletitle{What makes a good code example?: A study of
  programming Q\&A in StackOverflow}. In \bibinfo{booktitle}{\emph{IEEE
  International Conference on Software Maintenance (ICSM)}}. IEEE.
\newblock


\bibitem[\protect\citeauthoryear{Ossher, Sajnani, and Lopes}{Ossher
  et~al\mbox{.}}{2011}]%
        {ossher11}
\bibfield{author}{\bibinfo{person}{Joel Ossher}, \bibinfo{person}{Hitesh
  Sajnani}, {and} \bibinfo{person}{Cristina Lopes}.}
  \bibinfo{year}{2011}\natexlab{}.
\newblock \showarticletitle{File cloning in open source java projects: The
  good, the bad, and the ugly}. In \bibinfo{booktitle}{\emph{IEEE International
  Conference on Software Maintenance (ICSM)}}. IEEE.
\newblock


\bibitem[\protect\citeauthoryear{PerlMonks}{PerlMonks}{[n.d.]a}]%
        {PerlMonksResponseToCox}
\bibfield{author}{\bibinfo{person}{PerlMonks}.}
  \bibinfo{year}{[n.d.]}\natexlab{a}.
\newblock \bibinfo{title}{Perl regexp matching is slow??}
\newblock \bibinfo{howpublished}{\url{https://perlmonks.org/?node_id=597262}}.
\newblock


\bibitem[\protect\citeauthoryear{PerlMonks}{PerlMonks}{[n.d.]b}]%
        {PerlRegexEngineSource_CacheCommentBlock}
\bibfield{author}{\bibinfo{person}{PerlMonks}.}
  \bibinfo{year}{[n.d.]}\natexlab{b}.
\newblock \bibinfo{title}{Snapshot of Perl 5 regex.c}.
\newblock
  \bibinfo{howpublished}{\url{https://web.archive.org/web/20190206210240/https://github.com/Perl/perl5/blob/blead/regexec.c}}.
\newblock


\bibitem[\protect\citeauthoryear{Phan, Nguyen, Nguyen, and Nguyen}{Phan
  et~al\mbox{.}}{2017}]%
        {Phan2017MiningAPIUsage}
\bibfield{author}{\bibinfo{person}{Hung~Dang Phan}, \bibinfo{person}{Anh~Tuan
  Nguyen}, \bibinfo{person}{Trong~Duc Nguyen}, {and} \bibinfo{person}{Tien~N.
  Nguyen}.} \bibinfo{year}{2017}\natexlab{}.
\newblock \showarticletitle{{Statistical migration of API usages}}. In
  \bibinfo{booktitle}{\emph{International Conference on Software Engineering
  Companion (ICSE-C 2017}}.
\newblock
\showISBNx{9781538615898}


\bibitem[\protect\citeauthoryear{Ponzanelli, Bacchelli, and Lanza}{Ponzanelli
  et~al\mbox{.}}{2013}]%
        {ponzanelli13}
\bibfield{author}{\bibinfo{person}{Luca Ponzanelli}, \bibinfo{person}{Alberto
  Bacchelli}, {and} \bibinfo{person}{Michele Lanza}.}
  \bibinfo{year}{2013}\natexlab{}.
\newblock \showarticletitle{Seahawk: Stack overflow in the ide}. In
  \bibinfo{booktitle}{\emph{International Conference on Software Engineering
  (ICSE)}}. IEEE Press.
\newblock


\bibitem[\protect\citeauthoryear{Ponzanelli, Bavota, Di~Penta, Oliveto, and
  Lanza}{Ponzanelli et~al\mbox{.}}{2014}]%
        {ponzanelli14}
\bibfield{author}{\bibinfo{person}{Luca Ponzanelli}, \bibinfo{person}{Gabriele
  Bavota}, \bibinfo{person}{Massimiliano Di~Penta}, \bibinfo{person}{Rocco
  Oliveto}, {and} \bibinfo{person}{Michele Lanza}.}
  \bibinfo{year}{2014}\natexlab{}.
\newblock \showarticletitle{Mining StackOverflow to turn the IDE into a
  self-confident programming prompter}. In \bibinfo{booktitle}{\emph{Working
  Conference on Mining Software Repositories (MSR)}}. ACM.
\newblock


\bibitem[\protect\citeauthoryear{Rahman, Bird, and Devanbu}{Rahman
  et~al\mbox{.}}{2012}]%
        {rahman12}
\bibfield{author}{\bibinfo{person}{Foyzur Rahman}, \bibinfo{person}{Christian
  Bird}, {and} \bibinfo{person}{Premkumar Devanbu}.}
  \bibinfo{year}{2012}\natexlab{}.
\newblock \showarticletitle{Clones: What is that smell?}
\newblock \bibinfo{journal}{\emph{Empirical Software Engineering}}
  \bibinfo{volume}{17}, \bibinfo{number}{4-5} (\bibinfo{year}{2012}),
  \bibinfo{pages}{503--530}.
\newblock


\bibitem[\protect\citeauthoryear{Rathnayake and Thielecke}{Rathnayake and
  Thielecke}{2014}]%
        {Rathnayake2014rxxr2}
\bibfield{author}{\bibinfo{person}{Asiri Rathnayake} {and}
  \bibinfo{person}{Hayo Thielecke}.} \bibinfo{year}{2014}\natexlab{}.
\newblock \bibinfo{booktitle}{\emph{{Static Analysis for Regular Expression
  Exponential Runtime via Substructural Logics}}}.
\newblock \bibinfo{type}{{T}echnical {R}eport}.
\newblock


\bibitem[\protect\citeauthoryear{Rattan, Bhatia, and Singh}{Rattan
  et~al\mbox{.}}{2013}]%
        {rattan13}
\bibfield{author}{\bibinfo{person}{Dhavleesh Rattan}, \bibinfo{person}{Rajesh
  Bhatia}, {and} \bibinfo{person}{Maninder Singh}.}
  \bibinfo{year}{2013}\natexlab{}.
\newblock \showarticletitle{Software clone detection: A systematic review}.
\newblock \bibinfo{journal}{\emph{Information and Software Technology}}
  \bibinfo{volume}{55}, \bibinfo{number}{7} (\bibinfo{year}{2013}),
  \bibinfo{pages}{1165--1199}.
\newblock


\bibitem[\protect\citeauthoryear{Ray, Kim, Person, and Rungta}{Ray
  et~al\mbox{.}}{2013}]%
        {Ray2013SemanticPortingProblems}
\bibfield{author}{\bibinfo{person}{Baishakhi Ray}, \bibinfo{person}{Miryung
  Kim}, \bibinfo{person}{Suzette Person}, {and} \bibinfo{person}{Neha Rungta}.}
  \bibinfo{year}{2013}\natexlab{}.
\newblock \showarticletitle{{Detecting and characterizing semantic
  inconsistencies in ported code}}. In \bibinfo{booktitle}{\emph{Automated
  Software Engineering (ASE)}}. \bibinfo{publisher}{IEEE}.
\newblock
\showISBNx{9781479902156}


\bibitem[\protect\citeauthoryear{Roichman and Weidman}{Roichman and
  Weidman}{2009}]%
        {Roichman2009ReDoS}
\bibfield{author}{\bibinfo{person}{Alex Roichman} {and} \bibinfo{person}{Adar
  Weidman}.} \bibinfo{year}{2009}\natexlab{}.
\newblock \showarticletitle{{VAC - ReDoS: Regular Expression Denial Of
  Service}}.
\newblock \bibinfo{journal}{\emph{Open Web Application Security Project
  (OWASP)}} (\bibinfo{year}{2009}).
\newblock


\bibitem[\protect\citeauthoryear{Roy, Cordy, and Koschke}{Roy
  et~al\mbox{.}}{2009}]%
        {roy09sci}
\bibfield{author}{\bibinfo{person}{Chanchal~K Roy}, \bibinfo{person}{James~R
  Cordy}, {and} \bibinfo{person}{Rainer Koschke}.}
  \bibinfo{year}{2009}\natexlab{}.
\newblock \showarticletitle{Comparison and evaluation of code clone detection
  techniques and tools: A qualitative approach}.
\newblock \bibinfo{journal}{\emph{Science of computer programming}}
  \bibinfo{volume}{74}, \bibinfo{number}{7} (\bibinfo{year}{2009}),
  \bibinfo{pages}{470--495}.
\newblock


\bibitem[\protect\citeauthoryear{Ruiz, Nagappan, Adams, and Hassan}{Ruiz
  et~al\mbox{.}}{2012}]%
        {ruiz12}
\bibfield{author}{\bibinfo{person}{Israel J~Mojica Ruiz},
  \bibinfo{person}{Meiyappan Nagappan}, \bibinfo{person}{Bram Adams}, {and}
  \bibinfo{person}{Ahmed~E Hassan}.} \bibinfo{year}{2012}\natexlab{}.
\newblock \showarticletitle{Understanding reuse in the android market}. In
  \bibinfo{booktitle}{\emph{IEEE International Conference on Program
  Comprehension (ICPC)}}. IEEE.
\newblock


\bibitem[\protect\citeauthoryear{Sadler, Lee, Lim, and Fullerton}{Sadler
  et~al\mbox{.}}{2010}]%
        {Sadler2010ResearchStrategy}
\bibfield{author}{\bibinfo{person}{Georgia~Robins Sadler},
  \bibinfo{person}{Hau-Chen Lee}, \bibinfo{person}{Rod Seung-Hwan Lim}, {and}
  \bibinfo{person}{Judith Fullerton}.} \bibinfo{year}{2010}\natexlab{}.
\newblock \showarticletitle{{Recruitment of hard-to-reach population subgroups
  via adaptations of the snowball sampling strategy}}.
\newblock \bibinfo{journal}{\emph{Nursing {\&} Health Sciences}}
  \bibinfo{volume}{12}, \bibinfo{number}{3} (\bibinfo{date}{9}
  \bibinfo{year}{2010}), \bibinfo{pages}{369--374}.
\newblock
\showISSN{14410745}


\bibitem[\protect\citeauthoryear{Saini, Farmahinifarahani, Lu, Baldi, and
  Lopes}{Saini et~al\mbox{.}}{2018a}]%
        {saini18fse}
\bibfield{author}{\bibinfo{person}{Vaibhav Saini}, \bibinfo{person}{Farima
  Farmahinifarahani}, \bibinfo{person}{Yadong Lu}, \bibinfo{person}{Pierre
  Baldi}, {and} \bibinfo{person}{Cristina~V Lopes}.}
  \bibinfo{year}{2018}\natexlab{a}.
\newblock \showarticletitle{Oreo: Detection of clones in the twilight zone}. In
  \bibinfo{booktitle}{\emph{European Software Engineering Conference and
  Symposium on the Foundations of Software Engineering (ESEC/FSE)}}. ACM.
\newblock


\bibitem[\protect\citeauthoryear{Saini, Sajnani, and Lopes}{Saini
  et~al\mbox{.}}{2018b}]%
        {saini18}
\bibfield{author}{\bibinfo{person}{Vaibhav Saini}, \bibinfo{person}{Hitesh
  Sajnani}, {and} \bibinfo{person}{Cristina Lopes}.}
  \bibinfo{year}{2018}\natexlab{b}.
\newblock \showarticletitle{Cloned and non-cloned Java methods: a comparative
  study}.
\newblock \bibinfo{journal}{\emph{Empirical Software Engineering}}
  (\bibinfo{year}{2018}), \bibinfo{pages}{1--47}.
\newblock


\bibitem[\protect\citeauthoryear{Sajnani, Saini, and Lopes}{Sajnani
  et~al\mbox{.}}{2014}]%
        {sajnani14}
\bibfield{author}{\bibinfo{person}{Hitesh Sajnani}, \bibinfo{person}{Vaibhav
  Saini}, {and} \bibinfo{person}{Cristina~V Lopes}.}
  \bibinfo{year}{2014}\natexlab{}.
\newblock \showarticletitle{A comparative study of bug patterns in java cloned
  and non-cloned code}. In \bibinfo{booktitle}{\emph{International Working
  Conference on Source Code Analysis and Manipulation (SCAM)}}. IEEE.
\newblock


\bibitem[\protect\citeauthoryear{Samet}{Samet}{1981}]%
        {Samet1981Lisp2Lisp}
\bibfield{author}{\bibinfo{person}{Hanan Samet}.}
  \bibinfo{year}{1981}\natexlab{}.
\newblock \showarticletitle{{Experience with software conversion}}.
\newblock \bibinfo{journal}{\emph{Software: Practice and Experience}}
  \bibinfo{volume}{11}, \bibinfo{number}{10} (\bibinfo{year}{1981}),
  \bibinfo{pages}{1053--1069}.
\newblock
\showISSN{1097024X}


\bibitem[\protect\citeauthoryear{Scacchi}{Scacchi}{2007}]%
        {scacchi07fse}
\bibfield{author}{\bibinfo{person}{Walt Scacchi}.}
  \bibinfo{year}{2007}\natexlab{}.
\newblock \showarticletitle{Free/open source software development: recent
  research results and emerging opportunities}. In
  \bibinfo{booktitle}{\emph{European Software Engineering Conference and
  Symposium on the Foundations of Software Engineering (ESEC/FSE)}}.
\newblock


\bibitem[\protect\citeauthoryear{Shen, Jiang, Xu, Yu, Ma, and Lu}{Shen
  et~al\mbox{.}}{2018}]%
        {Shen2018ReScueGeneticRegexChecker}
\bibfield{author}{\bibinfo{person}{Yuju Shen}, \bibinfo{person}{Yanyan Jiang},
  \bibinfo{person}{Chang Xu}, \bibinfo{person}{Ping Yu},
  \bibinfo{person}{Xiaoxing Ma}, {and} \bibinfo{person}{Jian Lu}.}
  \bibinfo{year}{2018}\natexlab{}.
\newblock \showarticletitle{{ReScue: Crafting Regular Expression DoS Attacks}}.
  In \bibinfo{booktitle}{\emph{Automated Software Engineering (ASE)}}.
\newblock
\showISBNx{9781450359375}


\bibitem[\protect\citeauthoryear{Siegmund, K{\"{a}}stner, Liebig, Apel, and
  Hanenberg}{Siegmund et~al\mbox{.}}{2014}]%
        {Siegmund2014MeasuringExperience}
\bibfield{author}{\bibinfo{person}{Janet Siegmund}, \bibinfo{person}{Christian
  K{\"{a}}stner}, \bibinfo{person}{Jörg Liebig}, \bibinfo{person}{Sven Apel},
  {and} \bibinfo{person}{Stefan Hanenberg}.} \bibinfo{year}{2014}\natexlab{}.
\newblock \showarticletitle{{Measuring and modeling programming experience}}.
\newblock \bibinfo{journal}{\emph{Empirical Software Engineering}}
  \bibinfo{volume}{19}, \bibinfo{number}{5} (\bibinfo{date}{10}
  \bibinfo{year}{2014}), \bibinfo{pages}{1299--1334}.
\newblock
\showISSN{1382-3256}


\bibitem[\protect\citeauthoryear{Sim, Clarke, and Holt}{Sim
  et~al\mbox{.}}{1998}]%
        {sim98}
\bibfield{author}{\bibinfo{person}{Susan~Elliott Sim},
  \bibinfo{person}{Charles~LA Clarke}, {and} \bibinfo{person}{Richard~C Holt}.}
  \bibinfo{year}{1998}\natexlab{}.
\newblock \showarticletitle{Archetypal source code searches: A survey of
  software developers and maintainers}. In
  \bibinfo{booktitle}{\emph{International Workshop on Program Comprehension
  (IWPC)}}. IEEE.
\newblock


\bibitem[\protect\citeauthoryear{Sipser}{Sipser}{2006}]%
        {Sipser2006AutomataTextbook}
\bibfield{author}{\bibinfo{person}{Michael Sipser}.}
  \bibinfo{year}{2006}\natexlab{}.
\newblock \bibinfo{booktitle}{\emph{Introduction to the Theory of
  Computation}}. Vol.~\bibinfo{volume}{2}.
\newblock \bibinfo{publisher}{Thomson Course Technology Boston}.
\newblock


\bibitem[\protect\citeauthoryear{Spencer}{Spencer}{1994}]%
        {Spencer1994RegexEngine}
\bibfield{author}{\bibinfo{person}{Henry Spencer}.}
  \bibinfo{year}{1994}\natexlab{}.
\newblock \showarticletitle{{A regular-expression matcher}}.
\newblock In \bibinfo{booktitle}{\emph{Software solutions in C}}.
  \bibinfo{pages}{35--71}.
\newblock


\bibitem[\protect\citeauthoryear{Staicu and Pradel}{Staicu and Pradel}{2018}]%
        {Staicu2018REDOS}
\bibfield{author}{\bibinfo{person}{Cristian-Alexandru Staicu} {and}
  \bibinfo{person}{Michael Pradel}.} \bibinfo{year}{2018}\natexlab{}.
\newblock \showarticletitle{{Freezing the Web: A Study of ReDoS Vulnerabilities
  in JavaScript-based Web Servers}}. In \bibinfo{booktitle}{\emph{USENIX
  Security Symposium (USENIX Security)}}.
\newblock


\bibitem[\protect\citeauthoryear{Terekhov}{Terekhov}{2001}]%
        {Terekhov2001LanguageConversion}
\bibfield{author}{\bibinfo{person}{A.A. Terekhov}.}
  \bibinfo{year}{2001}\natexlab{}.
\newblock \showarticletitle{{Automating language conversion: a case study (an
  extended abstract)}}.
\newblock \bibinfo{journal}{\emph{IEEE International Conference on Software
  Maintenance (ICSM)}} (\bibinfo{year}{2001}), \bibinfo{pages}{654--658}.
\newblock
\showISBNx{0-7695-1189-9}
\showISSN{1063-6773}


\bibitem[\protect\citeauthoryear{Terekhov and Verhoef}{Terekhov and
  Verhoef}{2000}]%
        {Terekhov2000LangConversion}
\bibfield{author}{\bibinfo{person}{Andrey~A. Terekhov} {and}
  \bibinfo{person}{Chris Verhoef}.} \bibinfo{year}{2000}\natexlab{}.
\newblock \showarticletitle{{The realities of language conversions}}.
\newblock \bibinfo{journal}{\emph{IEEE Software}} \bibinfo{volume}{17},
  \bibinfo{number}{6} (\bibinfo{year}{2000}), \bibinfo{pages}{111--124}.
\newblock
\showISSN{07407459}


\bibitem[\protect\citeauthoryear{Thompson}{Thompson}{1968}]%
        {Thompson1968LinearRegexAlgorithm}
\bibfield{author}{\bibinfo{person}{Ken Thompson}.}
  \bibinfo{year}{1968}\natexlab{}.
\newblock \showarticletitle{{Regular Expression Search Algorithm}}.
\newblock \bibinfo{journal}{\emph{Communications of the ACM (CACM)}}
  (\bibinfo{year}{1968}).
\newblock


\bibitem[\protect\citeauthoryear{Treude and Robillard}{Treude and
  Robillard}{2017}]%
        {treude17icsme}
\bibfield{author}{\bibinfo{person}{Christoph Treude} {and}
  \bibinfo{person}{Martin~P Robillard}.} \bibinfo{year}{2017}\natexlab{}.
\newblock \showarticletitle{Understanding stack overflow code fragments}. In
  \bibinfo{booktitle}{\emph{IEEE International Conference on Software
  Maintenance and Evolution (ICSME)}}. IEEE.
\newblock


\bibitem[\protect\citeauthoryear{Truskett}{Truskett}{[n.d.]}]%
        {PerlRegexDocs2}
\bibfield{author}{\bibinfo{person}{Iain Truskett}.}
  \bibinfo{year}{[n.d.]}\natexlab{}.
\newblock \bibinfo{title}{Perl Regular Expressions Reference - Perl}.
\newblock
  \bibinfo{howpublished}{\url{https://perldoc.perl.org/5.22.0/perlreref.html}}.
\newblock


\bibitem[\protect\citeauthoryear{Van Der~Merwe, Weideman, and Berglund}{Van
  Der~Merwe et~al\mbox{.}}{2017}]%
        {VanDerMerwe2017EvilRegexesHarmless}
\bibfield{author}{\bibinfo{person}{Brink Van Der~Merwe},
  \bibinfo{person}{Nicolaas Weideman}, {and} \bibinfo{person}{Martin
  Berglund}.} \bibinfo{year}{2017}\natexlab{}.
\newblock \showarticletitle{{Turning Evil Regexes Harmless}}. In
  \bibinfo{booktitle}{\emph{Proceedings of the South African Institute of
  Computer Scientists and Information Technologists (SAICSIT)}}.
\newblock
\showISBNx{9781450352505}


\bibitem[\protect\citeauthoryear{Vasilescu, Filkov, and Serebrenik}{Vasilescu
  et~al\mbox{.}}{2013}]%
        {vasilescu13}
\bibfield{author}{\bibinfo{person}{Bogdan Vasilescu}, \bibinfo{person}{Vladimir
  Filkov}, {and} \bibinfo{person}{Alexander Serebrenik}.}
  \bibinfo{year}{2013}\natexlab{}.
\newblock \showarticletitle{Stackoverflow and github: Associations between
  software development and crowdsourced knowledge}. In
  \bibinfo{booktitle}{\emph{2013 International Conference on Social
  Computing}}. IEEE, \bibinfo{pages}{188--195}.
\newblock


\bibitem[\protect\citeauthoryear{Veanes, De~Halleux, and Tillmann}{Veanes
  et~al\mbox{.}}{2010}]%
        {Veanes2010Rex}
\bibfield{author}{\bibinfo{person}{Margus Veanes}, \bibinfo{person}{Peli
  De~Halleux}, {and} \bibinfo{person}{Nikolai Tillmann}.}
  \bibinfo{year}{2010}\natexlab{}.
\newblock \showarticletitle{{Rex: Symbolic regular expression explorer}}.
\newblock \bibinfo{journal}{\emph{International Conference on Software Testing,
  Verification and Validation (ICST)}} (\bibinfo{year}{2010}).
\newblock
\showISBNx{9780769539904}
\showISSN{2159-4848}


\bibitem[\protect\citeauthoryear{Wang, Bai, and Stolee}{Wang
  et~al\mbox{.}}{2019}]%
        {Wang2019ExploringEvolution}
\bibfield{author}{\bibinfo{person}{Peipei Wang}, \bibinfo{person}{Gina~R Bai},
  {and} \bibinfo{person}{Kathryn~T Stolee}.} \bibinfo{year}{2019}\natexlab{}.
\newblock \showarticletitle{{Exploring Regular Expression Evolution}}. In
  \bibinfo{booktitle}{\emph{Software Analysis, Evolution, and Reengineering
  (SANER)}}.
\newblock


\bibitem[\protect\citeauthoryear{Wang and Stolee}{Wang and Stolee}{2018}]%
        {Wang2018RegexTestCoverage}
\bibfield{author}{\bibinfo{person}{Peipei Wang} {and}
  \bibinfo{person}{Kathryn~T Stolee}.} \bibinfo{year}{2018}\natexlab{}.
\newblock \showarticletitle{{How well are regular expressions tested in the
  wild?}}. In \bibinfo{booktitle}{\emph{Foundations of Software Engineering
  (FSE)}}.
\newblock
\showISBNx{9781450355735}


\bibitem[\protect\citeauthoryear{Waters}{Waters}{1988}]%
        {Waters1988ProgramTranslation}
\bibfield{author}{\bibinfo{person}{Richard Waters}.}
  \bibinfo{year}{1988}\natexlab{}.
\newblock \showarticletitle{{Program translation via abstraction and
  reimplementation - Software Engineering}}.
\newblock \bibinfo{journal}{\emph{IEEE Transactions on Software Engineering}}
  \bibinfo{volume}{14}, \bibinfo{number}{8} (\bibinfo{year}{1988}).
\newblock


\bibitem[\protect\citeauthoryear{Weideman, van~der Merwe, Berglund, and
  Watson}{Weideman et~al\mbox{.}}{2016}]%
        {Weideman2016REDOSAmbiguity}
\bibfield{author}{\bibinfo{person}{Nicolaas Weideman}, \bibinfo{person}{Brink
  van~der Merwe}, \bibinfo{person}{Martin Berglund}, {and}
  \bibinfo{person}{Bruce Watson}.} \bibinfo{year}{2016}\natexlab{}.
\newblock \showarticletitle{{Analyzing matching time behavior of backtracking
  regular expression matchers by using ambiguity of NFA}}. In
  \bibinfo{booktitle}{\emph{Lecture Notes in Computer Science (including
  subseries Lecture Notes in Artificial Intelligence and Lecture Notes in
  Bioinformatics)}}, Vol.~\bibinfo{volume}{9705}. \bibinfo{pages}{322--334}.
\newblock
\showISBNx{9783319409450}
\showISSN{16113349}


\bibitem[\protect\citeauthoryear{{Wikipedia contributors}}{{Wikipedia
  contributors}}{2018}]%
        {regexwikipedia}
\bibfield{author}{\bibinfo{person}{{Wikipedia contributors}}.}
  \bibinfo{year}{2018}\natexlab{}.
\newblock \bibinfo{title}{Regular expression --- {Wikipedia}{,} The Free
  Encyclopedia}.
\newblock
  \bibinfo{howpublished}{\url{https://web.archive.org/web/20180920152821/https://en.wikipedia.org/w/index.php?title=Regular_expression}}.
\newblock


\bibitem[\protect\citeauthoryear{Wong, Yang, and Tan}{Wong
  et~al\mbox{.}}{2013}]%
        {wong13}
\bibfield{author}{\bibinfo{person}{Edmund Wong}, \bibinfo{person}{Jinqiu Yang},
  {and} \bibinfo{person}{Lin Tan}.} \bibinfo{year}{2013}\natexlab{}.
\newblock \showarticletitle{Autocomment: Mining question and answer sites for
  automatic comment generation}. In \bibinfo{booktitle}{\emph{Automated
  Software Engineering (ASE)}}. IEEE.
\newblock


\bibitem[\protect\citeauthoryear{Wustholz, Olivo, Heule, and Dillig}{Wustholz
  et~al\mbox{.}}{2017}]%
        {Wustholz2017Rexploiter}
\bibfield{author}{\bibinfo{person}{Valentin Wustholz}, \bibinfo{person}{Oswaldo
  Olivo}, \bibinfo{person}{Marijn J~H Heule}, {and} \bibinfo{person}{Isil
  Dillig}.} \bibinfo{year}{2017}\natexlab{}.
\newblock \showarticletitle{{Static Detection of DoS Vulnerabilities in
  Programs that use Regular Expressions}}. In
  \bibinfo{booktitle}{\emph{International Conference on Tools and Algorithms
  for the Construction and Analysis of Systems (TACAS)}}.
\newblock


\bibitem[\protect\citeauthoryear{Yang, Hussain, and Lopes}{Yang
  et~al\mbox{.}}{2016}]%
        {yang16}
\bibfield{author}{\bibinfo{person}{Di Yang}, \bibinfo{person}{Aftab Hussain},
  {and} \bibinfo{person}{Cristina~Videira Lopes}.}
  \bibinfo{year}{2016}\natexlab{}.
\newblock \showarticletitle{From query to usable code: an analysis of stack
  overflow code snippets}. In \bibinfo{booktitle}{\emph{Mining Software
  Repositories (MSR)}}. ACM, \bibinfo{pages}{391--402}.
\newblock


\bibitem[\protect\citeauthoryear{Yang, Martins, Saini, and Lopes}{Yang
  et~al\mbox{.}}{2017}]%
        {Yang2017SOGitHubSnippets}
\bibfield{author}{\bibinfo{person}{Di Yang}, \bibinfo{person}{Pedro Martins},
  \bibinfo{person}{Vaibhav Saini}, {and} \bibinfo{person}{Cristina Lopes}.}
  \bibinfo{year}{2017}\natexlab{}.
\newblock \showarticletitle{{Stack Overflow in Github: Any Snippets There?}}.
  In \bibinfo{booktitle}{\emph{IEEE International Working Conference on Mining
  Software Repositories (MSR)}}.
\newblock
\showISBNx{9781538615447}
\showISSN{21601860}


\bibitem[\protect\citeauthoryear{Zhang, Upadhyaya, Reinhardt, Rajan, and
  Kim}{Zhang et~al\mbox{.}}{2018}]%
        {Zhang2018AreSOCodeExamplesReliable}
\bibfield{author}{\bibinfo{person}{Tianyi Zhang}, \bibinfo{person}{Ganesha
  Upadhyaya}, \bibinfo{person}{Anastasia Reinhardt}, \bibinfo{person}{Hridesh
  Rajan}, {and} \bibinfo{person}{Miryung Kim}.}
  \bibinfo{year}{2018}\natexlab{}.
\newblock \showarticletitle{{Are Online Code Examples Reliable? An Empirical
  Study of API Misuse on Stack Overflow}}. In
  \bibinfo{booktitle}{\emph{International Conference on Software Engineering
  (ICSE)}}.
\newblock
\showISBNx{9781450356381}
\showISSN{02705257}


\bibitem[\protect\citeauthoryear{Zhong, Thummalapenta, Xie, Zhang, and
  Wang}{Zhong et~al\mbox{.}}{2010}]%
        {Zhong2010APIMapping}
\bibfield{author}{\bibinfo{person}{Hao Zhong}, \bibinfo{person}{Suresh
  Thummalapenta}, \bibinfo{person}{Tao Xie}, \bibinfo{person}{Lu Zhang}, {and}
  \bibinfo{person}{Qing Wang}.} \bibinfo{year}{2010}\natexlab{}.
\newblock \showarticletitle{{Mining API mapping for language migration}}. In
  \bibinfo{booktitle}{\emph{International Conference on Software Engineering}}.
\newblock
\showISBNx{9781605587196}
\showISSN{0270-5257}


\end{thebibliography}
\bibliographystyle{ACM-Reference-Format}

\end{document}